\documentclass[a4paper,fleqn,usenatbib]{mnras}

\usepackage{mathptmx}
\usepackage{txfonts}

\usepackage{graphicx}	
\usepackage{amsmath}	
\usepackage{amssymb}	

\usepackage{ulem}
\usepackage{multicol}

\title[From cusps to cores: a stochastic model]{From cusps to cores: a stochastic model}

\author[Amr A. El-Zant, Jonathan Freundlich and Fran\c coise Combes]{
 Amr A. El-Zant$^{1}$\thanks{E-mail: amr.elzant@bue.edu.eg}, Jonathan Freundlich$^{2,3}$ and  Fran\c coise Combes$^{2,3}$
\\
$^{1}$Centre for Theoretical Physics, The British University in Egypt, Sherouk City 11837, Cairo, Egypt\\
$^{2}$Coll\`ege de France, PSL Research University, F-75005, Paris, France\\
$^{3}$LERMA, Observatoire de Paris, CNRS, Sorbonne Universit\'es, UPMC Univ. Paris 06, F-75014, Paris, France
}

\date{Accepted XXX. Received YYY; in original form ZZZ}

\pubyear{2016}

\begin{document}
\label{firstpage}
\pagerange{\pageref{firstpage}--\pageref{lastpage}}

\maketitle

\begin{abstract}
The cold dark matter model  of structure formation faces apparent problems on galactic scales. 
Several threads point to excessive halo concentration, including central densities that rise too steeply with 
decreasing radius.   Yet, 
random fluctuations in the gaseous component can 'heat' the centres of haloes, 
decreasing their densities. We present a theoretical model deriving this effect from first principles: 
stochastic variations in the gas density are converted into potential fluctuations that act on the dark matter; 
the associated force correlation function is calculated and the corresponding 
stochastic equation solved. Assuming a power law spectrum of fluctuations with maximal and minimal cutoff scales, we 
derive the velocity dispersion imparted to the halo particles and the relevant relaxation time. 
We further perform numerical simulations, with fluctuations realised 
as a Gaussian random field, which confirm the formation of a core within a timescale comparable to that 
derived analytically. Non-radial collective modes enhance the energy transport process that erases the cusp, though the parametrisations of the analytical model persist.  
 In our model, the dominant contribution to the dynamical coupling driving the cusp-core transformation comes from the largest scale fluctuations. Yet, the efficiency of the transformation is independent of the value of the largest scale and depends weakly (linearly) on the power law exponent;
it effectively depends on two parameters: the gas mass fraction and the normalisation of the power spectrum.  This suggests that cusp-core transformations observed in hydrodynamic simulations of galaxy formation may be understood and parametrised in simple terms, the  physical and numerical complexities of the various implementations notwithstanding.
\end{abstract}






\section{Introduction}

Shortly after it was shown that simulated haloes within the  cold dark matter 
structure formation scenario display a singular central density profile up to the resolution radius \citep{Dubinski1991, Warren1992}, 
it was suggested that these might be in tension with observations of dark matter dominated galaxies
\citep{Flores1994, Moore1994}.
In most cases, finite density `cores' are  
favoured over singular `cusps'; and  there is, in general, simply too much mass
in the central regions of simulated CDM haloes for these to simultaneously fit the inner and outer rotation curves 
of dark matter dominated galaxies \citep[e.g., ][]{Weinberg2013}. This `cusp/core'  
conundrum seems particularly severe in dwarf galaxies \citep[e.g., ][]{Adams2014, Oh2015}. It is also probed in low surface brightness galaxies \citep[e.g., ][]{McGaugh1998a, Kuzio2011}, 
and may even be present in Milky Way satellites \citep{Goerdt2006, Walker2011} and the central galaxies of clusters \citep{Newman2015}.  Moreover, the high pattern speed of galactic bars in disk galaxies such as the Milky Way could also suggest the presence of a core rather than a cusp \citep{Debattista1998}.

In this context, the following questions arose:  
what precisely were the simulations 
predicting in terms of central slope of the density profile and mass contained in the central region 
of dark matter haloes? 
Were the profiles inferred from cosmological simulations a necessary theoretical
prediction of CDM cosmology?
And, finally, if actual discrepancies with observations do exist, how are these 
to be accounted for? 
Those questions have been addressed in numerous studies. 
Cosmological haloes were found to have an essentially universal density profile 
approximately characterised by the NFW formula \citep{nfw, Navarro1997}.  
The inner logarithmic slope is about $ -1$, though it  may flatten somewhat in the innermost
regions \citep{Stadel2009, Navarro2010}; 
the mass contained within the central region is determined by a concentration  
parameter, which correlates with the virial mass \citep{Bullock2001, Maccio2007, Klypin2011, Diemer2015}.
The combination of the form of the 
inner density profile and the mass concentration relation makes it difficult to fit the 
mass distribution inferred from observations. 
The second question pertains to theoretically 
understanding the origin of the profiles and the correlation between their parameters; despite 
much effort there is as yet no general theoretical model achieving this from first principles  
\citep[see ][ for a brief review]{Frenk2012}. Nevertheless, central cups do appear to be a generic product of
cold collapse \citep{Huss1999, Moore1999b, Shapiro2004, El-Zant2013}
and these cusps appear to be robust in the sense of being invariant under merging 
\citep{Kazantzidis2006, Boylan-Kolchin2004, El-Zant2008}.
The third question above may be related to other problems that collectively 
threaten the CDM paradigm; such as the `too big to fail' phenomenon \citep{Boylan-Kolchin2011}
which involves the excessive rotation speeds of galactic subhaloes and may 
be alleviated if those subhaloes are cored or with a shallow cusp 
\citep{Ogiya2015}.    
Proposed solutions can be broadly categorized
into those considering fundamental changes in the physics of the model and those concerned with the baryonic processes at stake during galaxy formation and evolution.
The first category comprises alternatives to cold, collisionless dark matter such as warm dark matter \citep[e.g., ][]{Colin2000, Bode2001, Schneider2012, Maccio2012b, Shao2013, Lovell2014, El-Zant2015}, self-interacting dark matter  \citep[e.g.,][]{Spergel2000,Burkert2000,Kochanek2000,Miralda2002, Peter2013, Zavala2013, Elbert2015}, and models that radically change the gravitational law \citep[e.g., ][]{Milgrom1983a, Gentile2011, Famaey2012}. Quantum effects are also sometimes invoked \citep[e.g., ][]{Goodman2000, Hu2000, Destri2013, Schive2014, Marsh2014, Chavanis2015}.

Given that the CDM paradigm only begins to face 
significant problems at precisely such scales when complex baryonic physics begins to 
play an important role, it is natural to inquire whether it is the central culprit behind erroneous theoretical predictions. 
It was for example realised early on that energy from supernovae may be sufficient for driving gas out of the potential wells of dwarf galaxies, 
the associated mass deficit resulting
in the expansion of the central halo region and the flattening of the density profile. 
More generally, many hydrodynamical simulations implementing stellar and AGN baryonic feedback processes in a cosmological context are able to reproduce cores \citep[e.g.,][]{Governato2010, Governato2012, Maccio2012, Martizzi2012, diCintio2014, Chan2015}. However, the complexity of such simulations obscures the physical mechanisms through which these processes affect the dark matter distribution.
These mechanisms 
normally invoke `heating' of the cold central density cusp 
through an irreversible process, such as dynamical friction from infalling 
clumps \citep{Zant2001, Zant2004, Tonini2006, RomanoDiaz2008, Goerdt2010, Cole2011, delPopolo2014, Nipoti2015}. 
Alternatively, repeated gravitational potential fluctuations 
induced by stellar winds, supernova explosions and active galactic nuclei (AGN)
could also dynamically heat the central halo 
\citep{Read2005, Mashchenko2006, Mashchenko2008, Peirani2008, Pontzen2012, Governato2012, Zolotov2012, Martizzi2013, Teyssier2013, Pontzen2014, Madau2014, Ogiya2014}.
%
Although the last mechanism  may seem  most closely related to the supernovae driven wind outflows 
discussed above, it is in principle more closely connected to the dynamical friction proposal, 
in the sense that it involves irreversible stochastic dynamics:
one may envisage the potential 
fluctuations leading to cusp-core transformation
as originating from stochastic density variations;  
the relevant `clumps' would be associated with fluctuation scales, as opposed 
to physically distinct  objects dissipating orbital energy {\it via} dynamical friction;
nevertheless, the basic physical mechanism  through which the energy 
is transferred to the dark matter is similar. For, as is the case  
in general with processes involving fluctuation and dissipation, 
fluctuations in a gravitational system can be approximated 
as stochastic processes described by power spectra and correlation functions, 
and they can be accompanied by dissipation in the form of dynamical 
friction \citep{Chandrasekhar1943, Nelson1999}.

{The purpose of this paper is to present and test a  
model for the case  when the fluctuations are driven by stellar winds, supernova explosions or AGN. 
The aim is  
to theoretically estimate the effect of such perturbations on the halo structure, given the shape of the 
density fluctuation power spectrum and its normalization. This should help in understanding 
the basic physics and dynamics of the process; to estimate the effect of 
potential fluctuations analytically or through simple simulations; and to interpret,
from first principles, complex hydrodynamical cosmological simulations, which differ in physical input and 
numerical implementation, and often on the inferred conclusion 
concerning the effectiveness of the process. 
At some level, the model incorporates scenarios whereby 
cusp-core transformation takes place due to potential 
variations arising from repeated outflows and inflows 
in the central region as the inferred mass variations, associated with the
density fluctuations, can be quite large in regions smaller than the largest 
fluctuation scales. In addition, it takes into account clumping and turbulent cascades 
that result in continuous mass and density fluctuation spectra.}
In Section~2 we outline the analytical model, solve it for power law spectra with 
cutoffs and derive an associated relaxation time, determining the timescale on which such fluctuations 
act to modify halo particle trajectories (details of the calculations 
are reproduced in the appendices). In Section~3 we test our model by evaluating 
the effect of the fluctuating field, with given power spectrum and normalization, 
on a live dark matter halo of the NFW form. Our conclusions are presented in Section~4.

\section{Dynamical relaxation spurred by stochastic density fields}
\label{section:analytics}

{
\subsection{Outline} 
\label{section:outline}
\subsubsection{Basic theoretical setup}

We envisage a two component system; a collisionless self gravitating 
system (primarily a dark matter halo) with smooth density distribution, 
which hosts a gaseous medium with density field exhibiting significant stochastic 
spatio-temporal variations in density. These can originate from stellar or AGN feedback; they   
lead to potential and force perturbations, which influence the motion of halo particles. 
These then deviate from their paths in the smooth potential within a  {\it relaxation time}. 
This is the time for the potential fluctuations to significantly affect particle trajectories; it is analogous to the  relaxation time in a stellar system, where the fluctuations due to 
point particle interactions can roughly be represented as white noise.
The relaxation time is  evaluated as follows. 

The density fluctuations are characterised by power spectra and 
associated correlation functions.  Once the power spectrum of density fluctuations is defined, 
the induced gravitational potential variations can be derived in a manner  
analogous to what  is done in calculations concerned with cosmological 
large scale structure. The force correlation function can then be evaluated from the potential fluctuations 
power spectrum, and from this the velocity variance imposed on halo 
particle trajectories by the force born of the density fluctuations. As in the standard calculation of stellar
dynamics, the velocity variance is then divided by the square of the average particle speed
and equated to unity to obtain the relaxation time, which is the characteristic time associated with the 
effect of potential fluctuations on the collisionless component. 

\subsubsection{Simplifying assumptions}
\label{sec:asump}

In order to render the model more tractable, and isolate the basic mechanism at work, 
we invoke some simplifying assumptions. 

We assume that the process we are interested
in occurs while the galaxy in question is gas rich. 
The collisionless component will therefore solely consist 
of a halo, assumed to initially  be in NFW form. 
Any stellar component present, being collisionless, would 
couple to the gas fluctuations as the dominant dark matter, 
though its initial distribution need not follow the halo of course. 
We therefore implicitly assume that this component's contribution 
to the mass distribution is small, especially as compared to the halo.
{This assumption will be especially justified if the star formation efficiency
is small, which is notably the case in low-surface-brightness and dwarf irregular galaxies \citep[e.g., ][]{vanderHulst1993, Schombert2001, Boissier2008, Wyder2009, Kennicutt2012}.
Such is also implied if hydrodynamical simulations invoking feedback 
are to simultaneously produce a halo core and match the stellar mass in 
dwarf galaxies \citep{Teyssier2013}. This assumption would seem even more justified if the energy input stems from 
AGN feedback.}
The gas mass fraction is taken to be independent 
of time; again, an implicit assumption here is that star formation is not efficient 
enough to induce a significant change in the gas mass fraction throughout the process. 

In this initial study, we confine ourselves to the case where
the gas is homogeneous on scales significantly larger than the largest 
fluctuation scales. Stellar and AGN feedback driven gaseous fluctuations are expected to be important in the central regions, and their dynamical effects, leading to core formation, significant within radii $\sim{r_s}$. 
We therefore assume that fluctuations are important only within a sphere of diameter $d$ around the centre.
Within this region, which is much larger than that bounded by $r_s$, 
we assume a gas fraction $f = {\rm M_{gas}/M_{DM}}$.
The gas mass fraction is a function of distance $l$ from the centre, thus in general $f= f(l)$.  
Since the gas is assumed to be homogeneous (barring fluctuations) and the dark matter is centrally 
concentrated, the gas mass fraction rapidly decreases with decreasing radius. 
{This means that if we assume, as we will,
a gas mass fraction of the order of the universal baryon fraction within $d/2$, we may be 
actually underestimating the gas fraction in the central regions
 (even though we do allow for gas condensation in the halo, as the region bounded by 
radius $d/2$ is assumed to be much smaller than the virial radius).  
However, the results presented are easily rescaled, as it will turn out that the gas mass fraction and
the normalisation of the density fluctuation power spectrum  enter multiplicatively in such a way that 
exactly the same results can be obtained by increasing the gas mass fraction and 
proportionally decreasing the assumed RMS density fluctuations.
Physically, a lower average central gas density accompanied by large fluctuations
may mimic repeated starburst/AGN driven outflows, leading to  
prolonged periods of small central gas mass fraction preceded by  
gaseous condensations and much larger 
than average densities.}

It is assumed that the affected collisionless matter distribution 
in the inner region of the halo remains near dynamical equilibrium. 
This naturally excludes haloes undergoing major mergers. 
However, cusps can reform during the merging process, and long lived cores
seem to emerge only after the epoch of rapid mass buildup is complete~\citep{Chan2015}.  
Moreover, as the gas mass is about an order
of magnitude smaller than the dark matter mass, the process 
of cusp-core transformation   {\it via} baryonic feedback 
should take place while the central halo remains in quasi-equilibrium. 
Indeed, it will turn out that the core cusp transformation takes place 
over many dynamical times for realistic choice of parameters.

In a system composed of dissipative gas that is repetitively driven by stellar winds, supernovae
or AGN energy input with stationary statistical properties, 
there should be a continuous power spectrum 
characterizing the fluctuations in the density field representing the transient gas clumps of different sizes. 
The general procedure outlined below is valid 
for any such power spectrum associated with a well defined
correlation function, as long as it decreases
sufficiently fast so that its integral converges. 
Nevertheless,  we assume that the 
power spectrum is a
power law with maximum and minimum cutoff.
This is motivated by theoretical considerations and 
observations of astrophysical fluids.

Fully turbulent media are  expected to display power law 
velocity spectra as fluctuations initiated at large scales would cascade into smaller scales down to the dissipation scale. 
If one associates the power law spectrum with  standard turbulence, the maximal scale is the 
energy driving scale, the standard (Kolmogorov) power law index is $5/3$. 
In compressible media the power spectrum of density fluctuations can 
approximately mimic that of their velocity counterparts, as seems to be the case in the cores of 
galaxy clusters \citep[e.g., ][]{Gaspari2013,Gaspari2014, Zhuravleva2015}.
It is also well established now that the interstellar medium (ISM) is highly inhomogeneous,
and characterized by supersonic velocities. The structure of the
ISM can be compared to a fractal structure, with a hierarchy of clumps with masses varying with 
the scale as a power law \citep[e.g.,][]{Larson1981, Falgarone1992, Elmegreen2002}. 
The slope of
the power law (or the fractal dimension) is between $1.5$ and  $2.0$ 
\citep{Chappell2001, Sanchez2005}.
The origin of the fractal could be self-gravity
\citep{Pfenniger1994, deVega1996}, with the minimum
and maximum scales being 10 AU and $\sim100~\rm pc$,  but turbulence and
magnetic fields have also been invoked \citep[e.g.,][]{Vazquez-Semadeni1997, Elmegreen1999, Padoan2004}.
Besides small-scale structures of the ISM, which might be bound by self-gravity,
there must exist kpc-scale structures, due to large-scale instabilities,
like spiral arms, or large kpc-scale clumps at high redshift
\citep[e.g.,][]{Noguchi1998,Bournaud2007,Elmegreen2009}.
In addition, the thermal and kinetic feedback due to starbursts and AGN
can create some transient kpc-size structures  
\citep{Stinson2006, DallaVecchia2008, Oppenheimer2008}.

Finally, in this section, we do not take into account the collective 
self gravitating response of the halo to the gas fluctuations.  
{It is also assumed, as in two body relaxation calculations,  
that  the velocity perturbations that result from the fluctuating force 
can be added to the unperturbed orbital motion of the halo particles. This excludes 
effects such as resonant coupling between the fluctuating force and the 
orbital motion of halo particles.} These effects are taken into account in
Section~3.

\subsubsection{Specific illustration}

Our general theoretical setup applies to any two component system 
as described above; and the calculations of the following sections
to any such system under the assumptions laid out in section~\ref{sec:asump}.
The analytical formulas can be rescaled to evaluate 
the effect of gaseous fluctuations on the dark matter halo (and stellar) profile for different 
collisionless matter distributions and gas mass fractions. 
 Nevertheless, for the sake 
of specific illustration, we will be focussing on the case of a small gas rich galaxy, 
assumed to be in early stages of evolution. The relevant parameters are given 
in Table~\ref{tab:parameters-halo}. We assume a gas mass fraction  of $0.17$ within the region where 
the fluctuations are considered important. However, we 
note that as the gas mass fraction and fluctuation levels 
enter multiplicatively in our calculations, these can be varied accordingly 
to get the same effect.
}

\subsection{The force correlation function}
\label{section:forcecor}

Let $\rho_0$ denote the average density of a fluid, representing galactic or
cluster gas that is driven by an energy source, causing large scale fluctuations within the fluid
over a volume $V = d^3$, the dimension $d$ being significantly larger 
than the largest fluctuation scale.   
The potential $\Phi$ and density contrast $\delta = \frac{\rho({\bf r})}{\rho_0} -1$ can be Fourier decomposed such that
\begin{equation}
\Phi  ({\bf r}) = \frac{V}{(2 \pi)^3}  \int \phi_{\bf k} e^{- i {\bf k . r}} d {\bf k},
\end{equation} 
and 
\begin{equation}
\delta ({\bf r}) = \frac{V}{(2 \pi)^3} \int \delta_{\bf k} e^{- i {\bf k . r}} d {\bf k}.
\end{equation} 
In this convention, physical and ${\bf k}$-space potential and 
density contrast have the same dimensions. If we assume the fluctuations define a stationary process so that ensemble averages
are time independent, the density fluctuation power spectrum is given by
\begin{equation}
	\mathcal{P}({\bf  k})   = V  \langle  |\delta_{\bf k}|^2  \rangle
\end{equation}
while the components $\phi_{\bf k}$ and $\delta_{\bf k}$ are related, via the Poisson equation
$ 
\nabla^2 \Phi= 4 \pi G \rho_0 \delta, 
$ 
through
\begin{equation}
\label{eq:phikk}
\phi_{\bf k} = -4 \pi G \rho_0 \delta_{\bf k} k^{-2}.  
\end{equation}

For a gaseous configuration that is isotropic on large scales, the force power spectrum is related to the potential fluctuations by 
\begin{equation}
\label{eq:pfk}
\mathcal{P}_F ({ k})  = V { k}^2 \langle |\phi_{ k}|^2  \rangle, 
\end{equation}
where $ k  = |{\bf k}|$. For a system that 
is furthermore homogeneous on large scales, the force correlation function, which is
the Fourier transform of the force power spectrum, is given by
\begin{equation}
\langle {\bf F} (0) . {\bf F} (r)\rangle =  \frac{V}{(2 \pi)^3} \int   k^2 \langle |\phi_k|^2\rangle  \frac{\sin (k r)}{k r} 4 \pi k^2 d k 
\end{equation}
We assume that the gaseous component is embedded in a dark matter halo and that the
fluctuations are present within some distance of about $d/2$ the centre 
of the halo mass distribution. This is the characteristic length scale within which the processes 
'stirring' the gas are significant and lead to fluctuations that display stationary statistical properties. 
One can then write
\begin{equation}
\langle {\bf F}(0) . {\bf F}(r) \rangle= \frac{d^3}{2 \pi^2 r} \int   k^3 \langle |\phi_k|^2 \rangle   \sin (k r) d k . 
\end{equation}
For power law density fluctuations
\begin{equation}
\label{eq:powerlaw}
\langle  |\delta_k|^2  \rangle = C k^{-n} , 
\end{equation}
the corresponding potential fluctuations are characterized by 
\begin{equation}
\langle  |\phi_k|^2  \rangle = (-4 \pi G \rho_0)^2 C k^{-4 -n}.    
\end{equation}
Consequently,
\begin{equation}
\langle {\bf F}(0) . {\bf F}(r) \rangle =   \frac{D}{r}     \int_{k_m}^{k_x }       \frac{\sin (k r)}{k^{n+1}} d k,  
\end{equation}
where
\begin{equation}
D = 8 (G \rho_0)^2 C d^3,
\label{eq:D}
\end{equation}
and where $k_m$ corresponds to the minimal fluctuation scale and $k_x$ to the maximal one. 
Assuming $k_x \gg k_m$, and  $n > 0$ the integral is evaluated in terms of incomplete Gamma functions 
to give
\begin{equation}
\langle {\bf F}(0) . {\bf F}(r) \rangle = - i k^{-n} \frac{D} {2 r}  (- i  k_m r)^n \Gamma (-n , - i k_m r) + C.C. ,  
\label{eq:F0Fr}
\end{equation}
where $C.C.$ refers to the complex conjugate. 
Note that for large $k_m r \gg 1$, that is for 
correlation between points separated by  distances much larger than the largest fluctuation scale $\lambda_{\rm max}/2\pi = 1/k_m$, 
\begin{equation}
\Gamma (-n , - i k_m r) \sim (- i k_m r)^{-n -1} e^{i k r}; 
\end{equation}
so that, in the diffusion limit we will 
be interested in below, 
	\begin{equation}
	\langle {\bf F}(0) . {\bf F}(r) \rangle
	\sim 
	\frac{D}{r^2}~ \frac{1}{k_{m}^{n+1}}~  \cos (k_{m}r). 
	\end{equation} 

\subsection{The velocity variance}
\label{section:velvar}

We are interested in the effect of the force fluctuations, born of the density variations in the gaseous field, on
the motions of the particles composing the surrounding halo. The velocity variance resulting from such 
effects can be evaluated as follows.  

Starting from the Newtonian equation ${d {\bf v}}/ {d t} =  {\bf F}$, and 
assuming that ${\bf F}$ is a random function with stationary statistical properties, one can multiply this equation by itself, take the ensemble 
average and change the time variables to obtain \citep[e.g., ][ and Appendix \ref{section:appendixB1}]{Osterbrock1952}
\begin{equation}
\langle (\Delta v)^2 \rangle = 2 \int_0^T (T - t)  \langle {\bf F}(0) . {\bf F}(t) \rangle d t. 
\label{eq:corrsp}
\end{equation}

Up to now, the correlation function was calculated in terms of spatial separation between two points.
This is interpreted in terms of an ensemble average over realisations of a fluctuating force field with 
stationary statistical properties.  The above equation on the other hand refers to the time 
correlation function along a halo particle trajectory. 
This temporal variation can be estimated by  considering  the motion of a
test halo particle with respect to the fluctuating gaseous field. 
The main contributions to its relative velocity will come from a mean flow 
(e.g., arising from its own orbital motion or fountain transporting the gas) 
as well as large scale random motions in the gas field.

The way the spatial statistical properties 
of the field are transported (or 'swept') into the temporal domain  due to such motions
has been extensively studied in the case of turbulent geophysical and atmospheric flows. 
If the spatial field properties are simply transported 
'frozen in'  {\it via} a bulk flow with average velocity $\langle v \rangle$ 
that is significantly larger than the velocities of the turbulent eddies,
the situation is similar to that invoked in the 
context of the classic \cite{Taylor1938} hypothesis. In this case, one has 
$\langle F(0) F(t) \rangle = \langle F(0) F(r =\langle v \rangle t) \rangle$.  
Since the velocities of large scale fluctuations in a fully turbulent medium are larger 
than those of the small scale eddies, the gaseous field can also be considered
to be  'randomly swept' with velocity $\langle u^2 \rangle^{1/2}$. This  corresponds to 
the random Taylor (or random sweeping) hypothesis \citep{Kraichnan1964, Tennekes1975}.
Theoretical, numerical and experimental studies, in the 
case of standard fully developed turbulence, suggest that in general 
the spatial statistical properties of the 
fluctuating field 
in the temporal domain can be related to those in the spatial one via a velocity 
$v_r = \sqrt{\langle   v \rangle^2  + \langle u^2 \rangle}$, such that the statistical 
properties in time at some given point 
are simply the spatial properties of the fluid transported with velocity $v_r$ through that point
\citep[e.g., ][]{Lvov1999, He2006, Zhao2009, He2011, Wilczek2012, Wilczek2014}. 
If this is the case then $\langle F(0) F(t) \rangle = \langle F(0) F(r = v_r t) \rangle$, where  
$ r =  v_r  t$ is the distance a test halo particle travels with respect to the 
fluctuating gas field during time $t$, both dues to its orbital motion and that of the field. 
Equation~(\ref{eq:corrsp}) can then be rewritten as 
\begin{equation}
\label{eq:intF0FR}
\langle (\Delta v)^2 \rangle = \frac{2}{v_r^2}  \int_0^{R = v_r T} (R - r)  \langle {\bf F}(0) . {\bf F}(r) \rangle d r, 
\end{equation}
which yields (Appendix~B2), when $k_x \gg k_m$,
\begin{equation}
\langle (\Delta v)^2 \rangle = \frac{Dk_m^{-n} R}{v_r^2}  \left(\frac{2}{n} {\rm Si}\left(k_{m} R\right)+T_1(k_{m} R)+T_2(k_{m} R)\right)
%
\end{equation}
where $\rm Si$ refers to the Sine integral and the transient terms are given by
\begin{equation}
\label{eq:T1}
T_1(k_{m} R) = \left(\frac{1}{n} - \frac{1}{n+1} \right) i (ik_{m}  R)^{n} \Gamma(-n,ik_{m}  R) + C.C.
\end{equation} 
and
\begin{equation} 
\label{eq:T2}
T_2 (k_{m} R) = \frac{2}{n+1} \frac{1}{k_{m}  R}   \left(\cos \left(k_{m}  R\right) - 1  \right).
\end{equation}
For $k_m R \gg 1$, these transient terms are much smaller than the first term inside the parenthesis, which converges to $\pi/n$.
Thus for large enough $R =    v_r t$, 
\begin{equation}
\langle (\Delta v)^2 \rangle =  \frac{ \pi D}{n  v_r}  \frac{T}{k_m^n}. 
\label{eq:variance}
\end{equation}
This is the formula we will use in estimating the effect of fluctuations in the 
gaseous medium on the trajectories of halo particles. It assumes that the 
particle moves a large enough distance $R  \gg k_m^{-1}$ with respect to the fluctuating field, i.e. much larger than the maximum
fluctuation scale $\lambda_{\rm max}/2 \pi$, 
either due to its own orbital motion or as a result of temporal variations in the field.
Such a regime corresponds to the diffusion limit, in which the halo particle is not ballistically displaced by steady forces  but instead 
undergoes a random walk initiated by the persistent density fluctuations.

For an unperturbed halo particle orbit of characteristic spatial extent $l$
the condition $R  \gg k_m^{-1}$ will be satisfied on the dynamical time
associated with the orbit provided that $l \gg \lambda_{\rm max}$. 
Since, for some orbits $l$ will be smaller than the largest fluctuation scales, 
this condition will not hold in general. However, as we assume that there
are no gaseous  inflows or outflows into or out of the region 
within radius $d/2$, within which the fluctuations are singificant, 
the largest scale gaseous 
motions will have typical scale $d/2 \gg \lambda_{\rm max}$. These gaseous motions 
are also expected to be more  effective in driving the approach to the 
diffusion limit, as they they are inherently random and non periodic. 
In the rest of this section we will thus assume that decorrelation occurs primarily 
due to the 'sweeping' of the flcutuating gas field by the largest scale gaseous motion. 
The condition that there are no inflows or outflows from $d/2$ suggests that 
the largest scale gaseous motion will be in energy equilibrium with the gravitational
field. The relevant velocity should therefore be $v_r \sim \frac{d/2}{t_D (d/2)}$, where $t_D (d/2)$ 
is the dynamical time at $d/2$.

Finally we note that, in the in the context of the sweeping hypotheses employed above, $v_r$ is assumed to be constant, independent of $k$.  If  $v_r$ is $k$-dependent,  an analogous calculation  would require 
that the transformation $r = v_r  t$  be already introduced in equation~(6), the correlation function 
will then depend on the form of $v_r(k)$. We do not pursue such cases here. Nevertheless, we 
will examine an example based on the Larson (1981) relation between the
velocity dispersion of a gaseous structure to its size in Section~3.
There, the orbital motion of halo particles is also explictly taken into account when simulating the 
motion of relative motion of the halo particles with respect to the fluctuating gaseous field.

\subsection{The relaxation time}
\label{section:trelax}

From Eq.~(\ref{eq:variance}), the ratio of the variance in velocity of the test particle 
produced by the fluctuating field to the square of the average orbital velocity of that particle is 
given by
\begin{equation}
\frac{\langle (\Delta v)^2 \rangle}{\langle v \rangle^2} =  \frac{ \pi D}{n  v_r  \langle v \rangle^2}  \frac{T}{k_m^n}.
\label{eq:relt} 
\end{equation}
As in the standard calculation of the two body relaxation  in stellar dynamics, we 
define the relaxation time as the time it takes for the left hand side of the 
above equation to become unity;
that is for the effect of fluctuations on the velocities to become numerically of the same order 
of the velocities in the smoothed out gravitational potential.  This gives
\begin{equation}
t_{\rm relax} =  \frac{n v_r \langle v \rangle^2 k_m^n}{\pi D}.
\end{equation}
Using Eq.~(\ref{eq:D}) for $D$ and noting that the constant determining the 
normalisation of the power spectrum can be  expressed in terms of the level of fluctuations 
at the maximum scale, such that $ C = k_m^n \langle \delta^2_{k_m}\rangle$, one gets
\begin{equation}
t_{\rm relax} = \frac{n v_r \langle v \rangle^2}{8 \pi (G \rho_0)^2\mathcal{P} (k_m)}, 
\end{equation}
where $\mathcal{P}  (k_m) =  \langle \delta^2_{k_m} \rangle  d^3$. 

Both velocities $v_r$ and $\langle v \rangle$ are determined by the gravitational field, which we assume to be dark matter dominated. Hence,
for some characteristic orbital scale $l$, $\langle v \rangle \sim l/t_D(l)$ with $t_D(l) \sim 1/ \sqrt{G \rho (< l)}$, where 
$\rho (< l)$ is the average density inside radius $l$.
Assuming that the main contribution to the motion of the particle velocity relative 
to the fluctuating gaseous field comes from large scale motions  with scale  $d/2$ in that field leads to
$v_r \sim d\sqrt{G \rho (< d/2)}/2$. Thus
\begin{equation}
t_{\rm relax} \approx \frac{n}{16 \pi \langle \delta^2_{k_m} \rangle}  f(l)^{-2}  \left(\frac{l}{d} \right)^2
\frac{\rho(< d/2)}{\rho(< l)}~ t_D (d/2)
\label{eq:trelax}
\end{equation}
where $f(l) = \rho_0/ \rho (<l)$ is the total gas fraction within radius $l$. 
Assuming a constant unperturbed gas density $\rho_0$, the relaxation time is a constant function of radius for a singular 
isothermal sphere of density $\rho(l) \propto l^{-2}$;
for shallower power law density cusps, it is an increasing function of radius. At $d/2$ it is generally given by
\begin{equation}
t_{\rm relax} (d/2)  \approx \frac{n}{64\pi \langle \delta^2_{k_m} \rangle }  f(d/2)^{-2} ~t_D (d/2).
\label{eq:scfree}
\end{equation}

The expressions  (\ref{eq:trelax}) and (\ref{eq:scfree}) for the relaxation time  are analogous to those obtained for two-body relaxation in $N$-body systems, for example in the case of stars or dark matter particles deflected by their successive interactions with one another \citep[e.g., ][]{Binney1987, Huang1993, ElZant2006}. Two-body relaxation is  modeled as a diffusion process due to random encounters between particles.
The associated relaxation time in the case of a system constituted by $N$ identical stars or dark matter particles can be expressed as 
	\begin{equation}
		t_{\rm relax} \sim 0.1 ~\frac{N}{\rm ln \Lambda} t_D, 
\label{eq:tbrelax}
	\end{equation}
	{\rm where $\Lambda = b_{\rm max}/b_{\rm min}$} is the ratio between the maximum and minimum impact parameters while $t_D$ is the dynamical time \citep{Binney1987}.  On the other hand, a random distribution 
of particles  leads to white noise ($n=0$ in the power spectrum). We do not 
deal with this special case in detail here; though we note that,  in such a system, the mass variance is expected to be proportional to $N$. Thus the variance associated with the relative density contrast $\langle \delta^2\rangle \propto 1/N$, and an expression analogous to~(\ref{eq:tbrelax}) can be deduced by using this result and imposing a maximum and 
minimum cutoff on a white noise power spectrum.  

If we assume the gas mass fraction inside radius $d/2$ 
is of order $0.17$, it can be seen from Eq. (\ref{eq:scfree}) that one needs $\langle \delta^2_{k_m} \rangle \sim 0.005$ in order to 
have a significant effect within a hundred dynamical 
times or so within the region where fluctuations are assumed to be present (for $ 2 \la n \la 3$).
As  $\langle \delta^2_{k_m} \rangle$ represents the contribution to the variance in density fluctuations
from the maximal fluctuation scale per $k$-space volume $\left(2 \pi/d\right)^3$, it depends explicitly 
on the scale of the region where the fluctuations are assumed to occur.
In the following we will evaluate the radial dependence of the relaxation time 
and relate its normalisation to descriptors of density fluctuations that do not bear this dependence, namely the power spectrum
(which is a measure of the contribution to the variance per unit $k$-space volume) and associated dimensionless spectrum and mass variance.

\section{Fluctuation levels and  their dynamical effects}

\subsection{Fluctuation levels  leading to relaxation on realistic timescale}

\begin{figure}
	\includegraphics[width=1.\columnwidth]{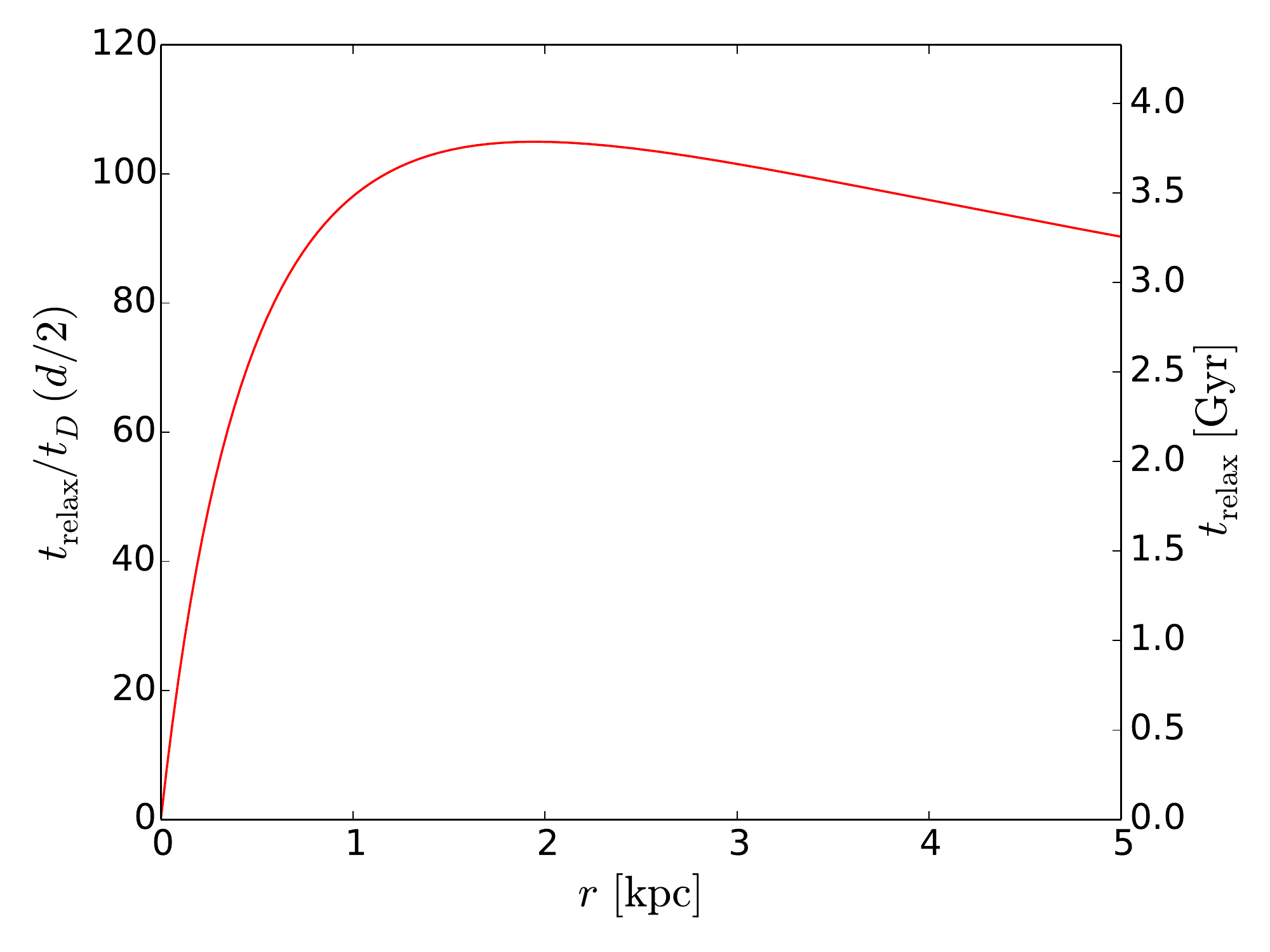}
    \caption{Evolution of the relaxation time as calculated from Eq.~(\ref{eq:trelax}) in the case of a fiducial NFW halo submitted to persistent density perturbations. The relaxation time is expressed in terms of the dynamical time $t_D(d/2)$ within $d/2 = 5~\rm kpc$ 
and in corresponding physical units}.
    \label{fig:trelax}
\end{figure}

In Fig.~\ref{fig:trelax} we plot the relaxation time deduced from equation~(\ref{eq:trelax})
for a galaxy with a dark matter halo of the NFW form. The halo is assumed to have a 
scale length $r_s = 0.9 ~{\rm kpc}$ and a total mass of $2.23 \times 10^{11} \rm M_\odot$ inside 
$R_{\rm vir} = 30~{\rm kpc}$. The gas mass fraction $f$ inside $d/2$ is 0.17. The power spectrum at 
maximum fluctuation scale is   $\mathcal{P} (k_m) = 4.6 ~{\rm kpc^3}$ 
(this corresponds to $\langle \delta_{k_m}^2 \rangle = 0.0046$ if $d = 10~{\rm kpc}$) and
the spectral index is taken as $n = 2.4$. These are the same parameters that we will use below in connection to what will be referred to as our fiducial simulation. The relaxation time is expressed in terms of the 
dynamical time within $d/2  = 5~\rm kpc$. Accordingly, our calculations suggest that, for the parameters chosen here, 
one should expect halo particles to be affected by potential fluctuations in the gas at all
radii $r \la r_s$ within a timescale of order of a hundred dynamical times.

\begin{figure}
	\includegraphics[width=1.\columnwidth]{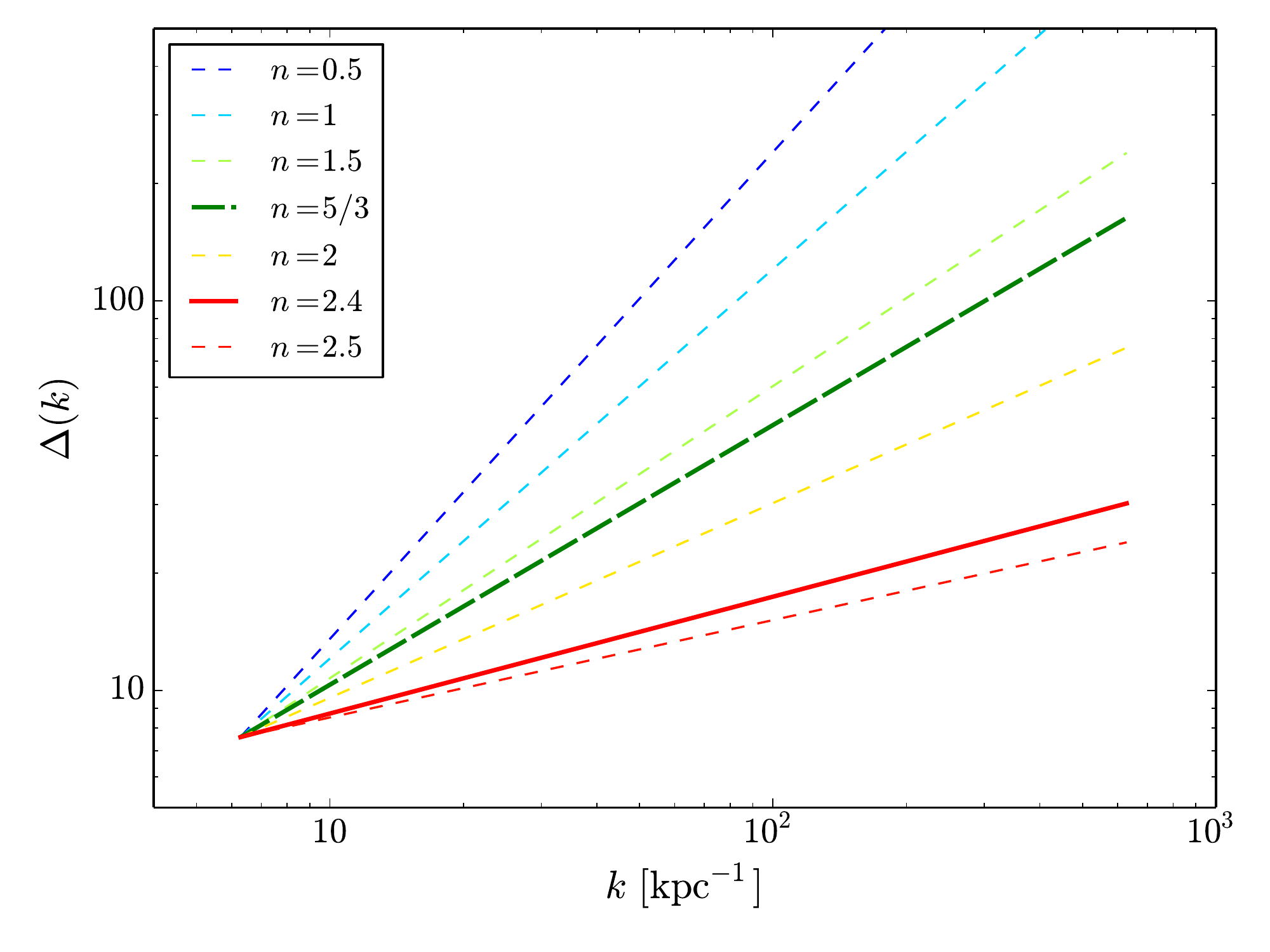}
    \caption{Dimensionless RMS fluctuations $\Delta(k)$ of the density contrast for a power law power spectrum as in Eq. \ref{eq:powerlaw}, with cutoff scales $\lambda_{\rm min} = 2\pi/k_x =  10 ~\rm pc$, $\lambda_{\rm max} = 2\pi/k_m = 1 ~\rm kpc$ and $\mathcal{P}(k_m) = 4.6~\rm kpc^3$. {The exponent $n$ increases from top to bottom.}}
    \label{fig:dimensionless}
\end{figure}

One would like
to quantify  the density fluctuation levels that are required in the gas in order to produce potential fluctuations 
leading to relaxation on that timescale. As the power spectrum has dimensions of volume it is not ideal for
this purpose. Instead, we estimate the  expected RMS fluctuations associated with the dimensionless power spectrum 
\begin{equation}
\Delta^2 (k) = \frac{k^3}{2 \pi^2} \mathcal{P}(k),
\end{equation}
which is a measure of the variance in density contrast $\delta$ per unit $\ln{k}$, 
which measures the contribution to the variance of in density fluctuations from logarithmic bins around wave number $k$. 
In Fig. \ref{fig:dimensionless} we plot $\Delta(k)$, for a power law power spectrum with fiducial cutoff scales, $\mathcal{P}(k_m) = 4.6^3~\rm kpc^3$ and different values of the exponent $n$.
The variance over all $k$ in density fluctuation contrast  is given by
\begin{equation}
\langle \delta^2 \rangle = \frac{1}{(2 \pi)^3)} \int_{k_m}^{k_x} \mathcal{P} (k) 4 \pi k^2 d k = \int_{k_m}^{k_x} \Delta^2(k) d\ln k, 
\end{equation}
hence
\begin{equation}
\langle \delta^2 \rangle \approx \frac{1}{2 \pi^2} k_m^n \mathcal{P}  (k_m) \frac{k_x^{3-n}}{3-n}
\end{equation}
for power law spectra with cutoff scales $k_x \gg k_m$ and $n < 3$.
As can be seen by plugging in typical values for $\mathcal{P} (k_m)$, $k_m$, $k_x$ and $n$
(say for maximum fluctuation scale $\lambda_{\rm max} = 2 \pi/k_m$ between of order 1~kpc, minimal scale 
0.01 to 0.1~kpc and $n = 1.5-2.5$) the inferred fluctuation levels are large. However, as we will see below, they appear compatible with those 
found in hydrodynamical  simulations where the effects discussed in this paper appear.

It is to be noted that 
although the variance in density contrast depends on $k_m$ and $k_x$, for power indexes considered 
here the {\it force}  fluctuations are dominated by the largest scales,  
so that fluctuations at $k_x$ are relatively unimportant; in addition, as we will see in section \ref{section:lmin-lmax}, in the 
diffusion limit, the dynamics is also independent of $k_m$. The crucial parameter therefore is the normalisation $\mathcal{P}(k_m)$. 
This is in line with what can be inferred from equation~(\ref{eq:trelax}).

\subsection{Mass fluctuations and power spectrum normalisation}

If one wishes to estimate the strength of density fluctuations in a realistic 
hydrodynamical simulation (or eventually possibly from observations), a natural
measure  is the variance of the average density on a particular scale $R$. 
If the random process giving rise to the density fluctuations is stationary 
the procedure involves measuring the  standard deviation of the mass over a sufficiently long timespan; or, for sufficiently small 
cells, the volume average of the mass and its square in a given snapshot (as in studies of large scale structure up to the limits imposed by 
cosmic variance). 
  
Theoretically, the variance can be evaluated, given a power spectrum, by filtering 
over different scales. Thus the variance  over a filtering scale $R$ is given by 
\citep[e.g., ][]{Martinez2002, Mo2010}
\begin{equation}
\label{eq:var}
\sigma_R^2 = \frac{1}{2 \pi^2}  \int_0^\infty W^2(k,R) \mathcal{P}(k) k^2 dk,
\end{equation}
where $W$ is the  Fourier transform of the window filtering function.  If the mechanism of core 
formation is indeed well modeled by the effects of  random Gaussian fluctuations
in the density field with a power law spectrum, then $\sigma_R$ derived from simulations
where cores are produced through potential fluctuations should be well fit 
by plugging in a power law $\mathcal{P} (k)$ into equation~(\ref{eq:var}); this constrains the normalisation 
$\mathcal{P} (k_m)$ and index $n$.

\begin{figure}
	\includegraphics[width=1.\columnwidth]{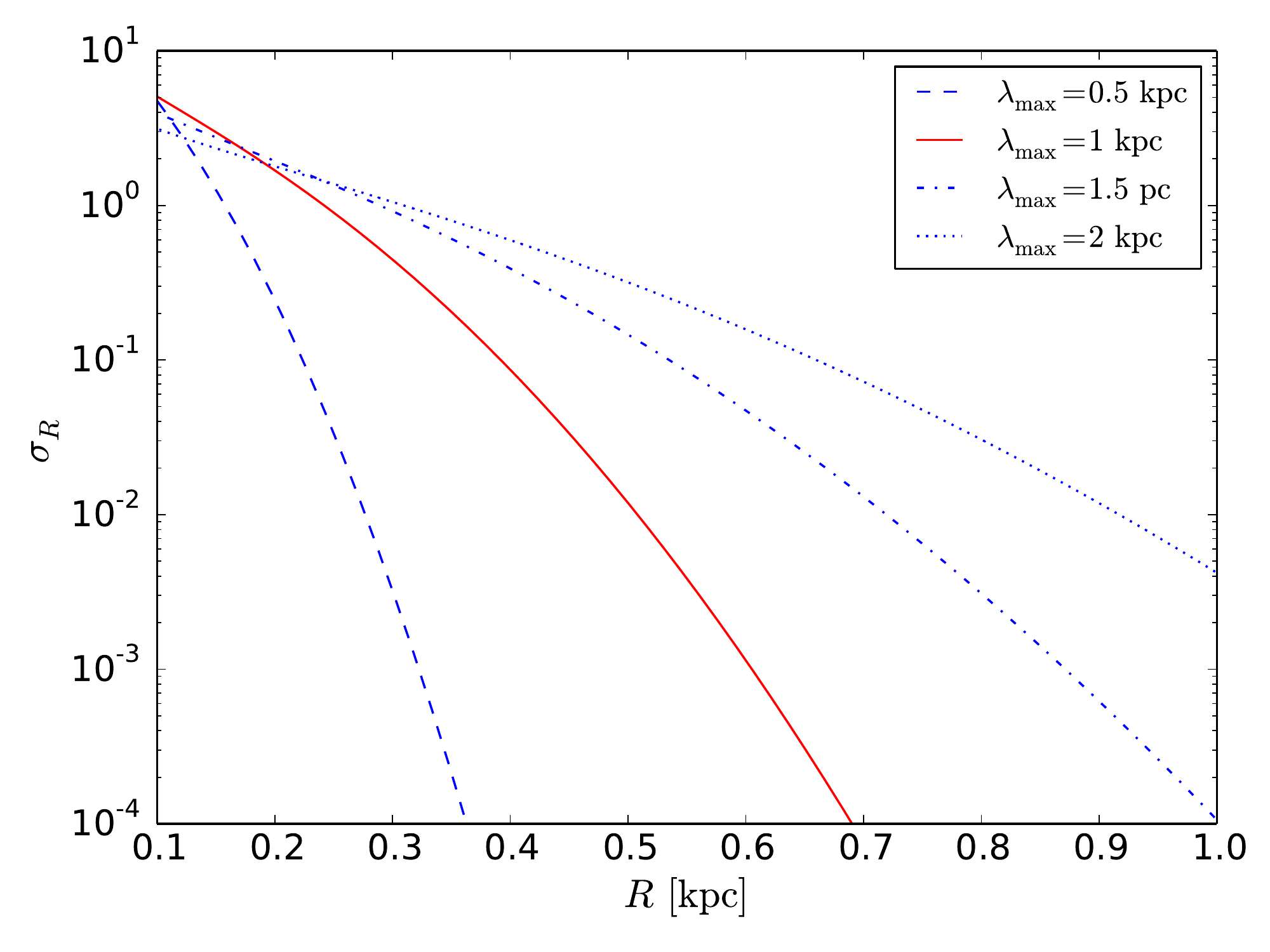}
    \caption{Relative RMS mass fluctuations averaged at different radii from Eq. \ref{eq:var} for a power law power spectrum. In this plot, we fix $\mathcal{P} (k_m) =  4.6 ~{\rm kpc^3}$, $n = 2.4$ and $\lambda_{\rm min} = 10 ~\rm pc$, but the curves are in fact largely independent of the power law exponent $n$ and of the minimum scale of the perturbations $\lambda_{\rm min}$. 
    \label{fig:sigmaR}
}
\end{figure}

As an example  we show in Fig.~\ref{fig:sigmaR} the RMS fluctuations as a function of radius enclosed using a Gaussian filter, 
$W (k, R) = e^{-k^2 R^2/2}$, for different values of the maximum fluctuation scale $\lambda_{\rm max}$. 
These  can be compared with RMS fluctuations in mass enclosed within radius $\sim R$ in hydrodynamical 
simulations. For example,~\cite{Teyssier2013} plot the variation in mass enclosed within different radii of an isolated 
dwarf galaxy simulated via the RAMSES code \citep{Teyssier2002}, with gas fluctuations driven by star formation.  
The  general level of fluctuations suggested by Fig.~\ref{fig:sigmaR} seems compatible with what can be inferred 
by eye from their Fig.~7. 
More quantitative, detailed comparison between our model and full hydrodynamical simulations
are left to a forthcoming study. Below, we will test our model in the more controlled context 
of a self gravitating halo with particles subjected to forces arising from the fluctuating density field.

{We note that, as the mass fluctuations in the central regions can be of order one 
or larger,  the assumed stochastic density fluctuations incorporate the effect of repeated 
rapid outflows (and subsequent inflows) that can be invoked as sources of non-adiabatic dynamics leading to core formation
\citep{Pontzen2012}. The fluctuations are rapid in our case, in the sense that their velocity $v_r$ 
is larger than the local orbital velocities in the central regions.}

\section{Numerical experiments}

In this section the fluctuations imposed on NFW halo particles are 
realised as  Gaussian random processes. The effect on the dark matter 
cusp is inferred. From the theoretical model described above we were 
able to estimate  the timescale on which the fluctuating force 
is expected to affect halo particle trajectories. This effect can be 
intuitively expected to drive the particles to higher energy levels  and thus 
lead to decrease in central density  and cusp-core transformation. 
We here show that this is indeed the case. 

{As opposed to the analytical calculations, the effect of the perturbations 
is not assumed to add to the motion in the smooth potential; instead, the 
equations of motion are solved with both contributions  (smooth potential plus 
fluctuating field) simultaneously included in the force term. Any
non-trivial  (e.g., resonant) coupling between the imposed fluctuating force and the orbital motion in the 
smoothed out potential is thus implicitly included.  The  self-gravity of the system of halo
particles, and hence its collective response to the potential fluctuations, is also taken into account.}
The diffusion limit is not assumed {\it a priori}.

\subsection{Code and initial conditions}

To evolve the dark matter distribution we use the self consistent field code (SCF)
of \cite{Hernquist1992}, which evaluates the density and potential 
via functional expansion suited for nearly spherical systems.  The setup is 
particularly powerful in capturing the interaction between stochastic processes
and large scale modes induced by self gravity \citep[e.g., ][]{Weinberg2013}; which turns out to be 
quite significant.

For the sake of specific illustration, we focus on the case of a small galaxy with an 
initial NFW halo with concentration parameter $c \approx 30$, scale length $r_s = 0.9 ~{\rm kpc}$
and mass ${M}_\mathrm{vir} = ~ 2.26 \times 10^{11} {\rm M_\odot}$ within radius $R_{\rm vir} = ~ c r_s$.
Given that the fluctuations in the gas are expected to be important primarily in the central region, where feedback is most 
effective, and that the dynamical effects of the fluctuations are expected to be significant mainly within radius $l \sim r_s \ll d/2$, 
we only apply the fluctuating force to particles within $l < d/2 =5 ~{\rm  kpc}$.
The gas mass fraction within this region  is  $f(d/2) = 0.17$, in line with the large gas mass fractions observed in 
high redshift galaxies \citep[e.g., ][]{Daddi2010, Tacconi2010, Tacconi2013, Forster2011}.
Table \ref{tab:parameters-halo} summarises the parameters describing the halo initial conditions for our fiducial run. 

						  		            \begin{table}
						  		            \centering
						  		            \caption{Parameters describing the halo initial conditions for our fiducial run.}						  		            
						  		             \label{tab:parameters-halo}
						  		            \begin{tabular}{lll}
						  		    	        \hline
						  		    	        \rule{-4pt}{3ex}
						  		    	        Dark matter halo mass & ${M}_\mathrm{vir}$ &   $2.26~10^{11}~\mathrm{M}_\odot$ \\
						  		    	        NFW cut-off radius & ${R}_\mathrm{vir}$ &  30 kpc\\
						  		    	        NFW characteristic radius & ${r}_{s}$ &  0.9 kpc\\
						  		    	        Gas fraction & ${f (d/2)}$ &  0.17\\
						  		     	        \hline
						  		             \end{tabular}
						  		             \normalsize
						  		             \end{table}

\subsection{Realisation of the power spectrum in terms of  Gaussian random field}

In this realisation, the density fluctuations are felt on the particle via a stochastic 
force. In line with the general setup presented in section~\ref{section:analytics},  the contribution of a density perturbation 
$\delta_{\bf k}$ to the stochastic force felt by a halo particle should be
				\begin{equation}
				\label{eq:force}
					{\bf F}_{\bf k} = - i {\bf k} \Phi_{\bf k} = 4\pi i G \rho_0 {\bf k}k^{-2} \delta_{{\bf k}}
				\end{equation}
		where $\rho_0$ corresponds to the homogeneous gas density that is assumed.
		As the force depends on the direction of ${\bf k}$, we consider a random direction ($\theta_k$, $\phi_k$) for each value of $k$, with $\theta_k \in \left[0,\pi\right]$ and $\phi_k \in \left[0,2\pi\right]$, so that 
		\begin{equation}
		    {\bf k} = k\left[\sin(\theta_k) \left(\cos(\phi_k) {\bf u}_x+\sin(\phi_k) {\bf u}_y\right) + \cos(\theta_k) {\bf u}_z\right],
		\end{equation}
	 and we introduce a random phase $\psi_k$ and the pulsation frequency associated to the density fluctuations $\omega(k)$. The force corresponding to mode $k$ felt by a halo particle situated at point ${\bf r}$ at time $t$ is consequently such that 
		\begin{equation}
		\label{eq:forcetot}
			{\bf F}_{\bf k} ({\bf r},t) \propto  {\bf k} ~ k^{-n/2-2} ~ \sin\left(\omega(k) t - {\bf k}. {\bf r} + \psi_k\right).
		\end{equation}

The force is rescaled {\it a posteriori} to match the assumed power spectrum normalisation, fixed through the choice of $\mathcal{P}(k_{m})$, as Eq.~(\ref{eq:phikk}) and hence (\ref{eq:forcetot}) only fix the relative values of $\phi_k$ while the absolute value depend on $V$. This 
is done using Eq.~(\ref{eq:F0Fr}), which yields the variance of the force resulting from the density fluctuations at all scales:
		\begin{equation}
		\label{eq:forcenorm}
			\langle F(0)^2\rangle = \frac{8\left(G\rho_0 \right)^2 \mathcal{P}(k_m)}{n-1} ~k_m ~ \left(1-\left(\frac{k_m}{k_{x}}\right)^{n-1}\right).
		\end{equation}
Again, we expect feedback processes and their associated dynamical effects on the halo to affect mostly the inner region, 
hence the fluctuating force ${\bf F}_{{\bf k}}$ is only applied when computing trajectories inside region of radius $d/2$.
Table~\ref{tab:parameters} summarizes the basic parameters used in deriving the perturbation force due to gaseous fluctuations 
for our fiducial run. 

				  		            \begin{table} 
				  		            \centering
				  		             \caption{Parameters describing the perturbations and their values for our fiducial run.
                                             The power spectrum normalisation corresponds to dimensionless power spectra and              mass variance shown on figures~\ref{fig:dimensionless} and~\ref{fig:sigmaR}. When the parameters of Table~\ref{tab:parameters-halo} are used, this normalisation results in the relaxation time shown in Fig.~\ref{fig:trelax}}
				  		             \label{tab:parameters}
				  		            \begin{tabular}{lll}
				  		    	        \hline
				  		    	        \rule{-4pt}{3ex}
				  		    	        Minimum scale  & $\lambda_\mathrm{min}= 2 \pi/k_x$ &  0.01 kpc\\
				  		    	        Maximum scale  &$\lambda_\mathrm{max} = 2 \pi/k_m$ &  1 kpc\\
				  		    	        Cutoff radius  &$r_{\rm cut}= d/2$ &  5 kpc\\
				  		    	        Power-law exponent &$n$ &  2.4 \\
				  		    	        Power spectrum at $k_m$ & $\mathcal{P} ({k_m})$ &  $4.6~{\rm kpc^3}$\\
				  		     	        \hline
				  		             \end{tabular}
				  		             \normalsize
				  		             \end{table}  

Finally, we need to choose the frequency of perturbation $\omega$.
For this purpose, we adopt two different approaches, each with its own 
set of simulations. 
The first set of simulations adopts the random sweeping approximation introduced in section~\ref{section:velvar}; the gas is assumed 
to be a fully turbulent medium with the smaller scales `swept' by the larger ones, which determines a common characteristic 
velocity independent of $k$. The characteristic timescale is 
$t_D (d/2) = 1/\sqrt{G \rho(<d/2)}$. The velocity associated to the Fourrier component then is 
$v_r = d/t_D(d/2)$ as in section \ref{section:trelax} and the frequency is given by $\omega(k) = v_r k$. 
Given our parameters and $d/2 = r_{\rm cut} = 5 ~\mathrm{kpc}$, we  have $v_{r} = 134 ~\mathrm{km.s}^{-1}$ for our fiducial simulation. This value lies in the velocity range observed for molecular outflows in nearby galaxies. Indeed, \cite{Cicone2014} report average outflow velocities ranging from 50 to 800 km.s$^{-1}$ in local ULIRGs and quasar-hosts with a median of about 200 km.s$^{-1}$. While outflow velocities can sometimes reach values close and above 1000 km.s$^{-1}$ \citep{Fischer2010, Sturm2011, Dasyra2012}, values of a few hundreds of km.s$^{-1}$ seem to be common \citep{Sakamoto2009, Combes2013}.

To examine the effect of $k$ dependence on $v_r$, in the  second set of simulations, we define $\omega(k)$ from Larson's relation, which relates the velocity dispersion of a gaseous structure 
to its size \citep{Larson1981, Solomon1987}. We can indeed expect the velocity associated to a density perturbation mode of size $\lambda$ to scale as its velocity dispersion $\sigma (\lambda)$. This latter quantity can be derived from Larson's relation \citep{Solomon1987}, 
		\begin{equation}
			\left(\frac{\sigma}{\mathrm{km.s}^{-1}}\right) \simeq 1.0 \left(\frac{\lambda}{\mathrm{pc}}\right)^{0.5}.
		\end{equation}
		Assuming that $v_r = d\omega/ dk$ equals $\sigma$, this empirical relation yields approximately 
			$\omega (k) \simeq 2  \sqrt{k}$,
		with $\omega$ in $(10 ~\mathrm{Myr})^{-1}$ and $k$ in kpc$^{-1}$,
		which also corresponds approximately to a characteristic time scale {$\tau =2\pi/\omega$} of 10 Myr for a kpc-sized structure. This time scale corresponds to the dynamical time within the gaseous structures and is comparable to the lifetimes of large star-forming molecular clouds, which are evaluated at a few tens of Myr \citep{Blitz1980, Larson1981, Elmegreen1991}. It is also of the same order of magnitude as the time needed for massive molecular outflows to expel the cold gas reservoir from a galaxy, as notably evaluated for Mrk 231 from IRAM Plateau-de-Bure CO(1-0) observations by \cite{Feruglio2010} or for a set of different galaxies by \cite{Sturm2011}. 
		
\subsection{Cusp flattening due to stochastic fluctuations: Strictly spherical case}
\label{collective}

At each time step during the SCF simulations, the density and the gravitational potential are approximated by a series of basis functions deriving from spherical harmonics which separate the radial and angular components. The expansion is truncated beyond radial and orbital `quantum' numbers $n_{max}$ and $l_{max}$, whose choice is crucial to capture the radial and angular structure of the halo without being dominated by the particle noise \citep[e.g.,][]{Weinberg1996, Meiron2014}. 

A cutoff $l_{max} = 0$, on the other hand,  corresponds to forcing strict spherical symmetry on the 
{self consistent} density and potential fields at each time step: density fluctuations are effectively smoothed out over the angular 
variables $\theta$ and $\phi$ and non-radial modes washed out. 
This eliminates the effect of non-radial global modes, but facillitates 
comparison with the analytical results in which the 
effect of collective self-gravitating response of the system is not taken into account at all. 
For, as in the case of standard two body relaxation, the relaxation time  is an estimate of the 
timescale over which individual trajectories are expected to perturbed due to the imposed fluctuating force, 
it does not take into account how energy may be transported and redistributed via global 
self-gravitating modes, which may affect the rate of evolution of the self-consistent mass distribution.
{To isolate the effect of non-radial collective modes we start by considering the case where 
strict spherical symmetry is maintained by the self consistent potential and density distribution. 
The imposed fluctuating force, however, is fully three dimensional as in the simulations with 
$l_{max} \ne 0$ discussed in the following subsections.}

\begin{figure}
						\centering
						\includegraphics[width=1.\columnwidth, trim = 0cm 0.3cm 0cm 0cm, clip]{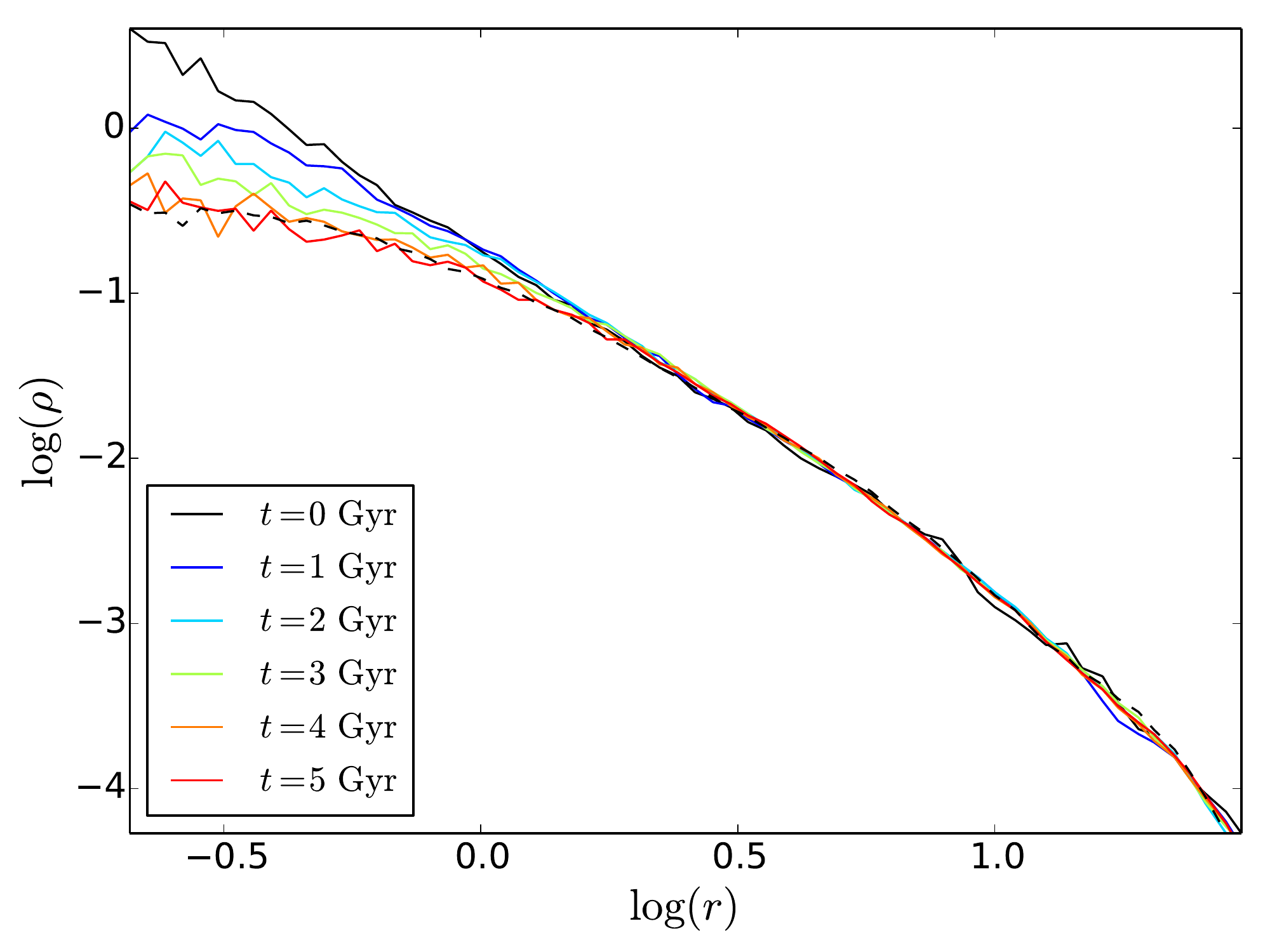}
						\caption{Evolution over 5 Gyr of the dark matter density profile with strict
spherically symmetry imposed on the halo density and potential at each time step ($l_{max} = 0$), {from an initially cuspy NFW profile to a flatter one}. The
halo is submitted to a fluctuating gravitational potential stemming from power-law density fluctuations as described in section \ref{section:analytics}; the parameters are those of Tables~\ref{tab:parameters-halo} and~\ref{tab:parameters}. 
 The radius $r$ is indicated in kpc while the density $\rho$ is in units of $2.26~10^{9}~\mathrm{M}_\odot/\mathrm{kpc}^3$. The pulsation frequency associated to each Fourier component was chosen as $\omega(k) = v_r k$, with a constant velocity $v_r = 134~\rm km.s^{-1}$. 
The rate of cusp-core transformation is in agreement with the analytical calculations (Fig.~\ref{fig:trelax}).
It is significantly slower however than the case when azimuthal modes are taken into account. For comparison, the black dashed line shows the averaged profile after 500~Myr for ten random realisations of a simulation including non-radial collective modes (as in Fig.~\ref{fig:fiducial}).} 
    \label{fig:lmax=0}
\end{figure}

The results are shown in  Fig.~\ref{fig:lmax=0}, for our fiducial parameters and constant $v_r$. 
 As can be seen the fluctuations imposed on the system of self gravitating halo particles does indeed produce a core from the initial cusp on the 
timescale predicted by the analytical calculations. 
This suggests that resonant coupling between the fluctuating force and the halo orbits is unimportant to the 
core formation process, and so are radial collective modes. As we will see below, on the other hand, 
azimuthal modes seem to significantly boost the core formation process. 

\subsection{Cusp flattening due to stochastic fluctuations: General case}

In this section we evaluate the effect of cusp flattening in the general case, without imposing 
strict spherical symmetry. After some trials, through which convergence of the results was verified,
we carry out simulations with $n_{max} = 10$ and $l_{max} = 4$. The results of
\cite{Vasiliev2013} suggest that this combination should be optimal, given the number of particles ($N = 240 000$).
In this context, we repeat the simulation of the previous subsection.
We also examine the effect of k-dependent speed $v_r$ of the gaseous field relative to dark matter particles, and the dependence of the process on maximal and minimal fluctuation scales and the power law exponent, as well as on the time resolution of the simulations. 

\subsubsection{Accelerated cusp-core transformation} 

\begin{figure}
\includegraphics[width=1.\columnwidth, trim = 0cm 0.3cm 0cm 0cm, clip]{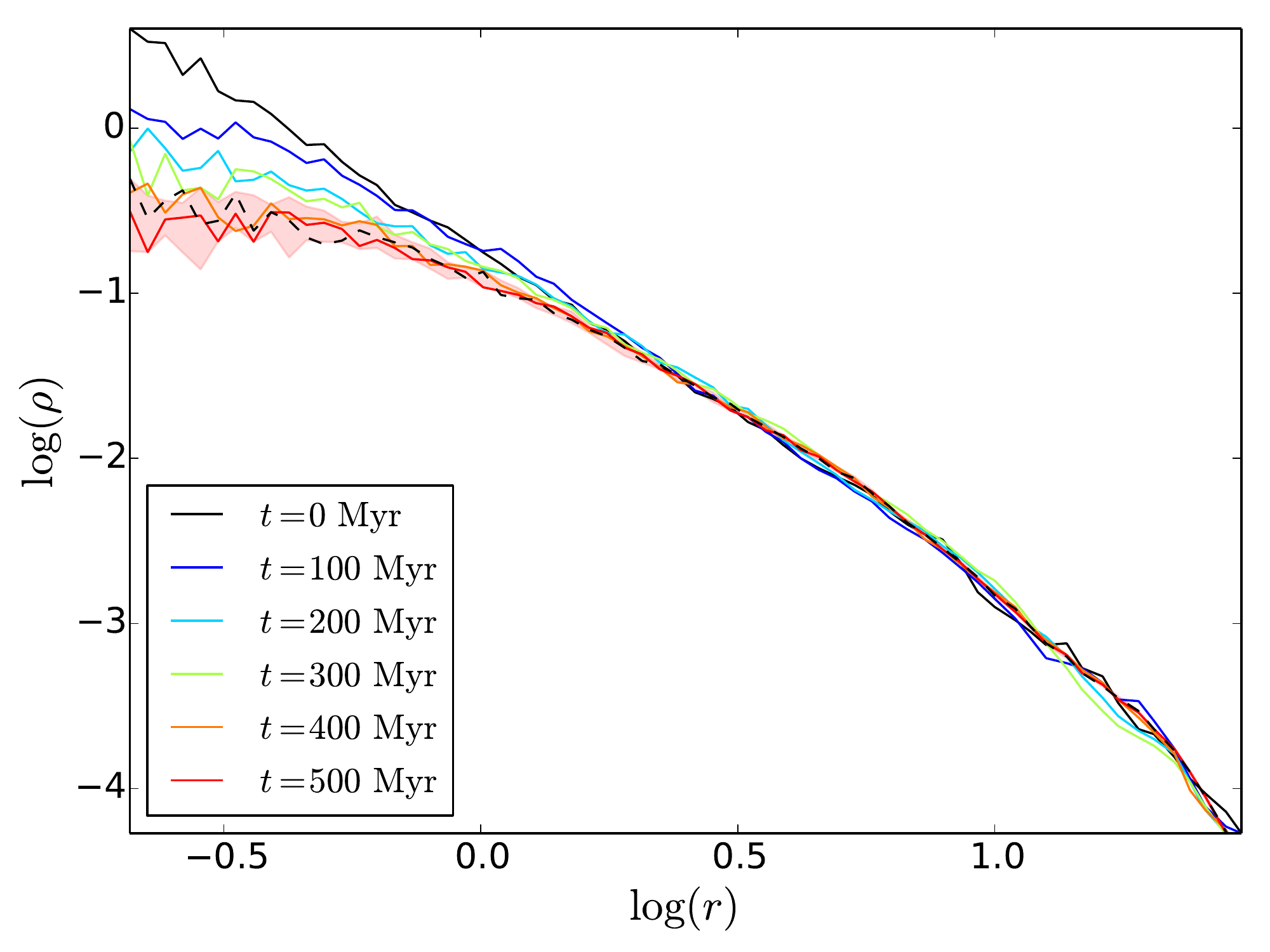}
\caption{Evolution of the dark matter density profile with parameters given in Tables~\ref{tab:parameters-halo} 
and~\ref{tab:parameters} and no strict spherical symmetry imposed. The solid lines correspond to the case when the pulsation frequency associated to each Fourier component was chosen as $\omega(k) = v_r k$, with a constant velocity $v_r = 134~\rm km.s^{-1}$. The shaded area highlights the scatter at $t =\rm 500~Myr$ between ten random realisations of the simulation. Alternatively, the dashed line displays the dark matter density profile after 500 Myr when the pulsation frequency is  defined from Larson's
relation as $\omega(k) = 2\sqrt{k}$: both approaches yield similar results.}
    \label{fig:fiducial}
\end{figure}

In Fig.~\ref{fig:fiducial} we show the evolution of initial cusped profile under the influence 
of a stochastic  force born  of density fluctuations as described in the previous subsection. This is done 
for a constant speed $v_r$ of the halo particles with respect  to the fluctuating field 
(as in the theoretical calculations of section \ref{section:analytics}), as well of $k$-dependent speed derived from 
Larson's relation. The parameters are those of the fiducial simulation (as summarized in Tables \ref{tab:parameters-halo} and \ref{tab:parameters}). 
 As can be seen, by the end of the simulation
(500 Myr), there is a significant effect at all radii within the initial NFW scale length $r_s =0.9~ {\rm kpc}$,
 and this effect is to erase the central cusp, transforming it into a nearly constant density core.
At larger radii, particle trajectories are also affected by 
the stochastic force, but the effect is similar at all radii (as the relaxation time flattens); the overall effect 
is suppressed and the shape of density at these radii remains largely unaltered.

Evidently, the core-cusp transformation is significantly faster here than in the case when strict spherical 
symmetry was imposed.  This phenomenon suggests that the azimuthal smoothing
suppresses  the energy redistribution within the halo and slows its collective response.
As the perturbation imposed on the halo particle trajectories is the same as in the case when spherical symmetry is enforced, the difference must stem from how the imparted energy is transported and redistributed within the halo, a process which can involve collective modes activated by self-gravity. 
That stochastic noise can excite global 'sloshing' modes, enhancing its overall effect, has been previously 
realised by \cite{Weinberg1998}. {Thus, while direct resonances between the imposed force 
and the dark matter orbits seem unimportant in 
our case, secondary resonances with such modes may act as to speed up the process of velocity dispersion equalisation in the cusp, and hence of its transformation into a core.}~\cite{Pontzen2015} also note that a  
triaxial halo submitted to time-dependent potential fluctuations flattens into a core within 1 Gyr while a similar spherical 
halo remains cuspy on such a timescale: asphericity seems to be a key ingredient for an efficient cusp-core transition.

\begin{figure}
	\includegraphics[width=1.\columnwidth, trim = 0cm 0.3cm 0cm 0cm, clip]{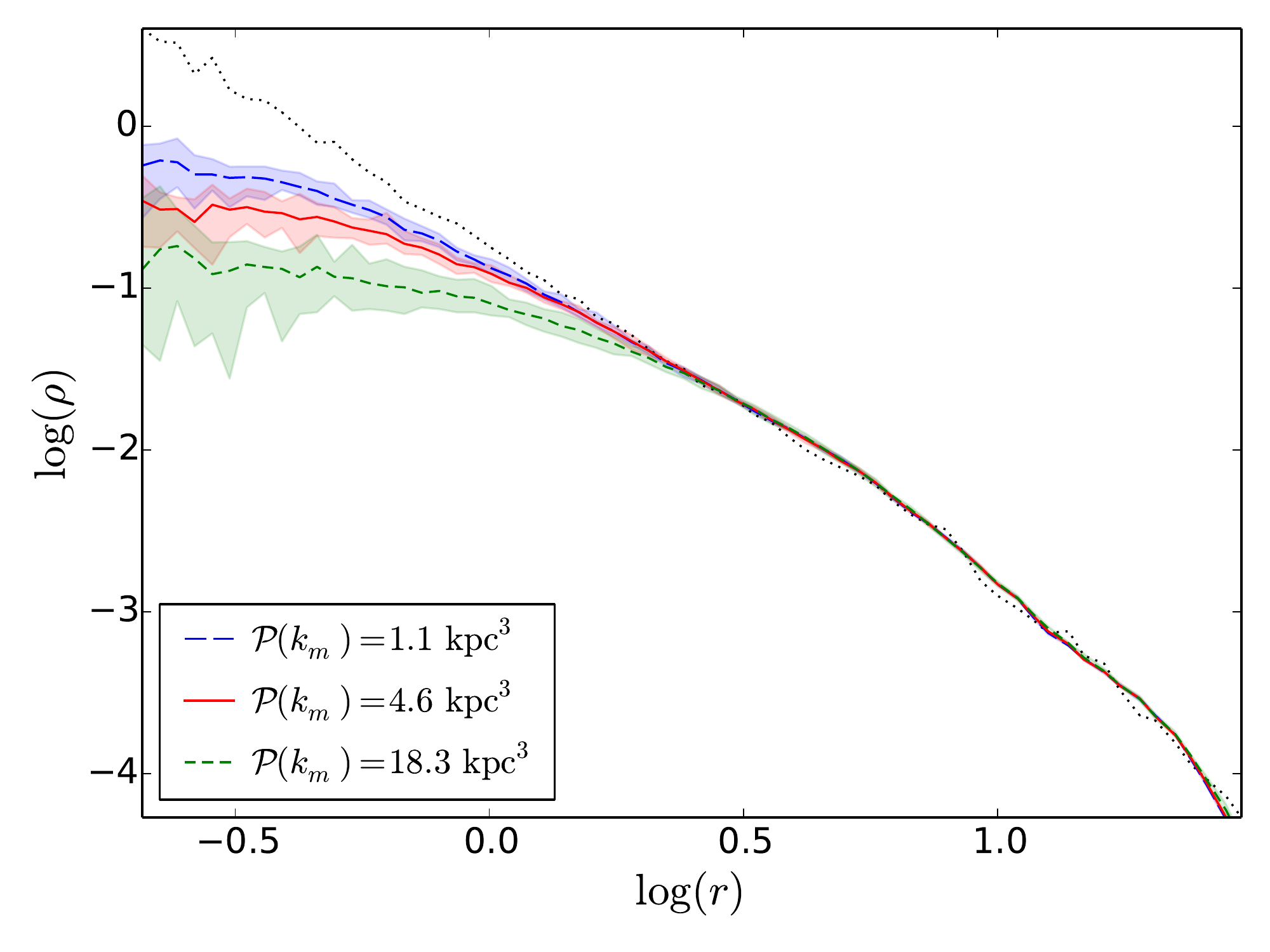}
    \caption{Evolution of the dark matter density profile after 500 Myr for different values of $\mathcal{P}(k_m)$ in the case of density fluctuations with constant speed of the turbulent flow with respect to the dark matter. The power spectrum normalisation grows by a factor 4 between two successive curves. 
    Each profile has been averaged over ten random realizations of the simulation; the shaded areas correspond to the span of these ten realizations. The dotted line corresponds to the initial profile; the units are as in the previous figures.
	}
    \label{fig:delta}

\end{figure}

\subsubsection{Power spectrum normalisation and gas mass fraction}

As may be expected, the normalisation the density fluctuation power spectrum, and hence 
of the imposed force (as defined by Eq.~\ref{eq:forcenorm}),
plays an important part in the magnitude of the ensuing effect. This is illustrated in 
Fig.~\ref{fig:delta}, where we show the effect of varying the values of the power spectrum normalisation $\mathcal{P}(k_m)$. 
To factor out the variations due to different random initial conditions we average 
the results over ten runs for each value and show the contours associated with these values.  

\begin{figure}
	\includegraphics[width=1.\columnwidth, trim = 0cm 0.3cm 0cm 0cm, clip]{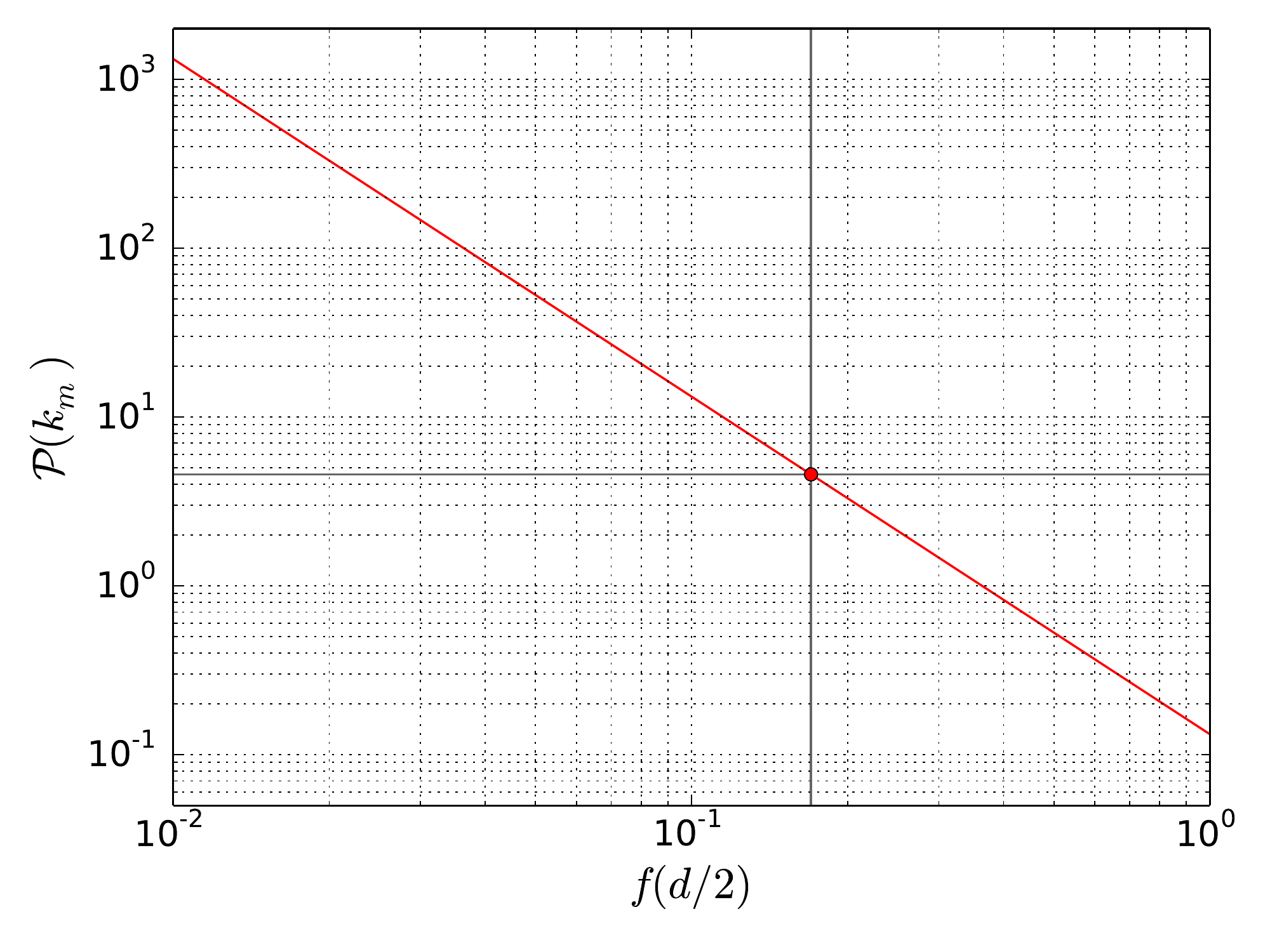}
    \caption{Values of the gas mass fraction $f(d/2)$ and of the power spectrum normalisation $\mathcal{P}(k_m)$ that keep the force normalisation as that assumed in the fiducial simulation, from Eq. (\ref{eq:forcenorm}). The minimum and maximum cutoff scales and the power law exponent are left unchanged. 
	}
    \label{fig:force-params}
\end{figure}

We note that since the gas mass fraction and the fluctuation levels enter multiplicatively in our formulation, 
we could have changed the gas mass fraction instead of the power spectrum normalisation to obtain analogous results.
Fig. \ref{fig:force-params} shows the gas mass fraction variation with power spectrum normalisation that keep the force normalisation at the same level
as that assumed in the fiducial simulation. {As already mentioned (section~\ref{section:outline}), by assuming 
a universal baryon fraction within a sphere of diameter $d$, we may be underestimating the 
gas mass fraction in the central regions. However the results are easilly rescaled, as exactly the same effect 
can be obtained for larger gas mass fraction by assuming smaller fluctuation levels.}

\subsubsection{Independence of the flattening on the maximum and minimum fluctuation scales}
\label{section:lmin-lmax}

The normalisation of the fluctuating force (Eq.~\ref{eq:forcenorm}) only weakly depends on the
minimum fluctuation scale $k_x$ as $k_x \ll k_m$, but does depend on the maximum 
fluctuation scale determined by $k_m$. Nevertheless, in the diffusion limit, 
the actual effect on particle trajectories, as determined by $\langle (\Delta v)^2 \rangle$, 
is not expected to depend on $k_m$ if the velocity of the perturbations relative to the halo particles 
$v_r$ is independent of $k$. 
This can be explained using the following heuristic
argument. 
In a diffusion process, particle trajectories are affected by small successive kicks, each of them associated to a velocity change $\Delta v \sim  F \Delta t$, where $\Delta t$ is the characteristic duration of the kick. If we assume a pulsation frequency $\omega = v_r k$, $\Delta t$ varies as $1/k$. 
The square of the kicks adds up linearly; such that, after a given time interval in which a test halo particle is subjected to $N$ kicks, the resulting velocity variance is $ \langle(\Delta v)^2 \rangle \sim N (\Delta v)^2$. However, since $\Delta t \propto 1/k$, the number of kicks during this time interval is proportional to $k$. Consequently,
$
			\langle(\Delta v)^2 \rangle \propto N~ F^2 ~\Delta t^2 \propto F^2/k.
$
As $\langle F(0)^2\rangle \propto k_{m}$ from Eq. (\ref{eq:forcetot}) in the limit $k_x \ll k_m $, the resulting velocity variance should be independent of $k_{m}$ even though this wavenumber determines the dominant scale of the perturbations.
		
\begin{figure}
						\centering
						\includegraphics[width=1.\columnwidth, trim = 0cm 1.85cm 0cm 0cm, clip]{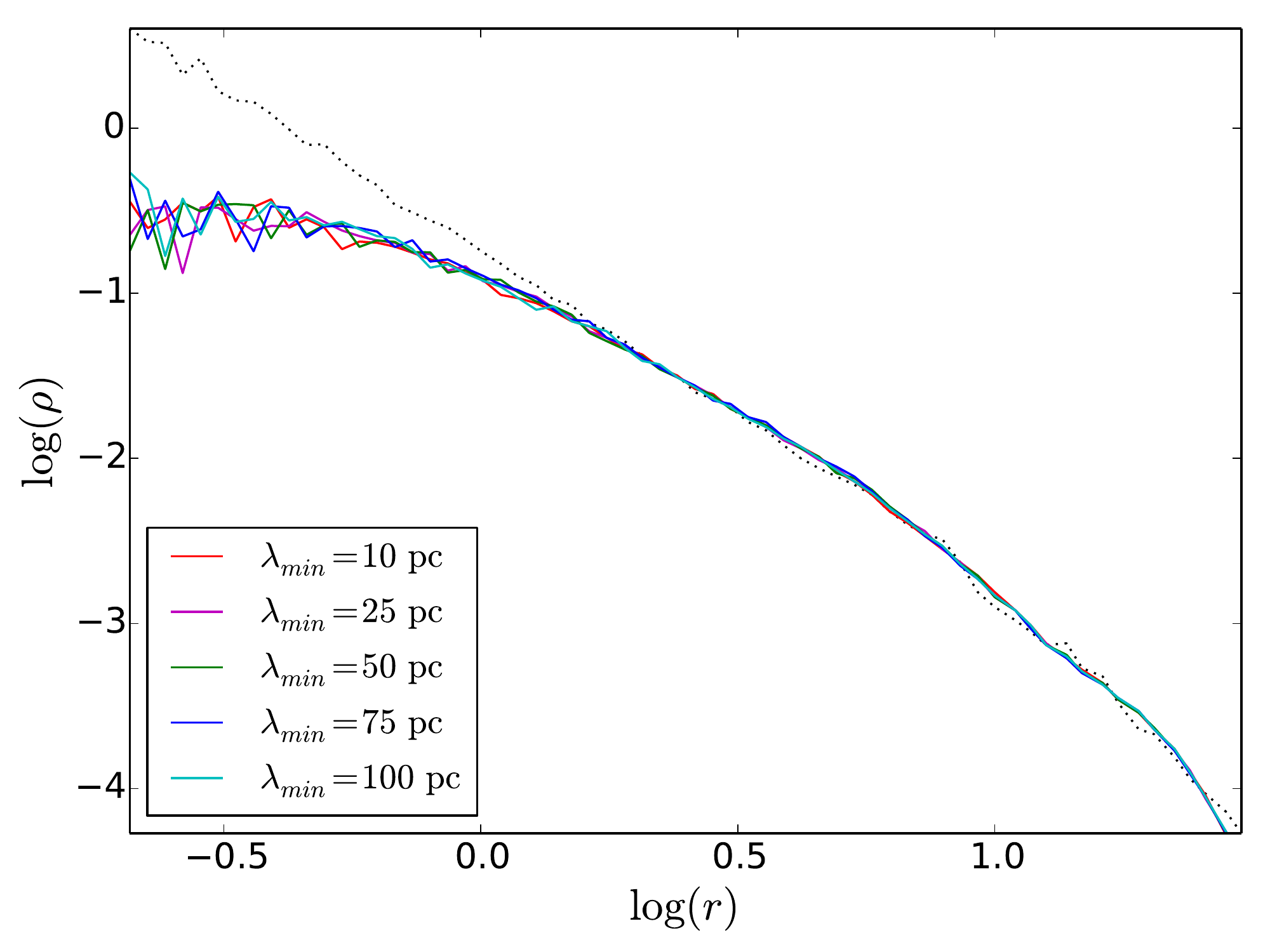}
						\includegraphics[width=1.\columnwidth, trim = 0cm 0.3cm 0cm 0cm, clip]{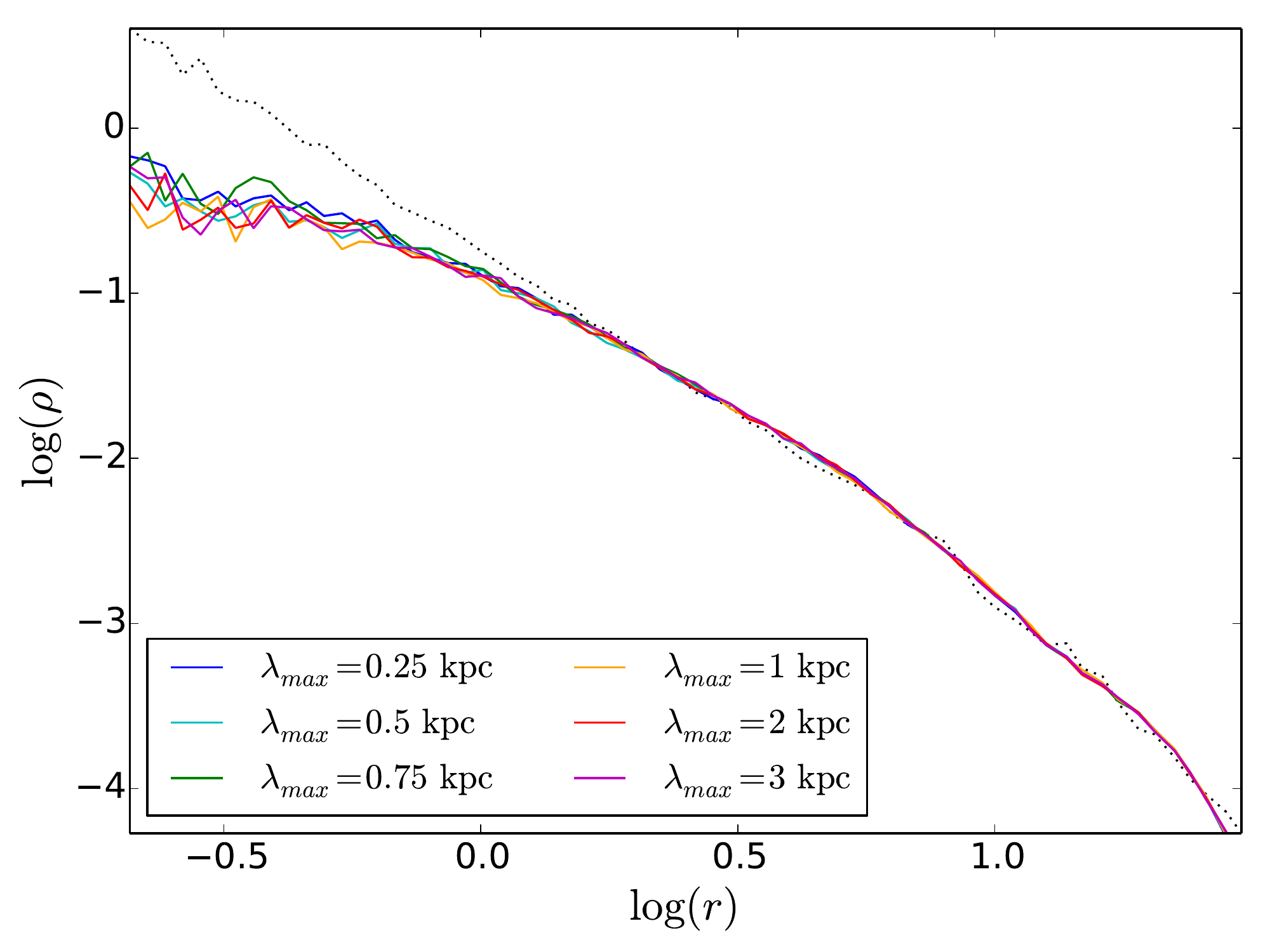}
						\caption{Dark matter density profiles after 500 Myr of perturbations for different values of the minimum (up) and maximum (down) scales of the perturbations. The parameters of the simulations correspond to the fiducial ones, with a constant speed of turbuleng flow $v_r$. The maximum scale of perturbations is  $\lambda_{\rm max}= \rm 1 ~kpc$ when varying $\lambda_{\rm min}$ (upper panel), and $\lambda_{\rm min } = 10 ~\rm pc$ when varying $\lambda_{\rm max}$ (lower panel). The dotted line corresponds to the initial profile; the units are as in the previous figures.} 
    \label{fig:lambda}
\end{figure}

Fig.~\ref{fig:lambda} shows that the effect of the fluctuations is indeed largely independent of 
the maximal and minimal fluctuation scales.  This is also expected from the analytical formula 
for the relaxation time (Eq.~(\ref{eq:trelax}).

\subsubsection{Comparison with a Kolmogorov exponent}

\begin{figure}
						\centering
						\includegraphics[width=1.\columnwidth, trim = 0cm 0.3cm 0cm 0cm, clip]{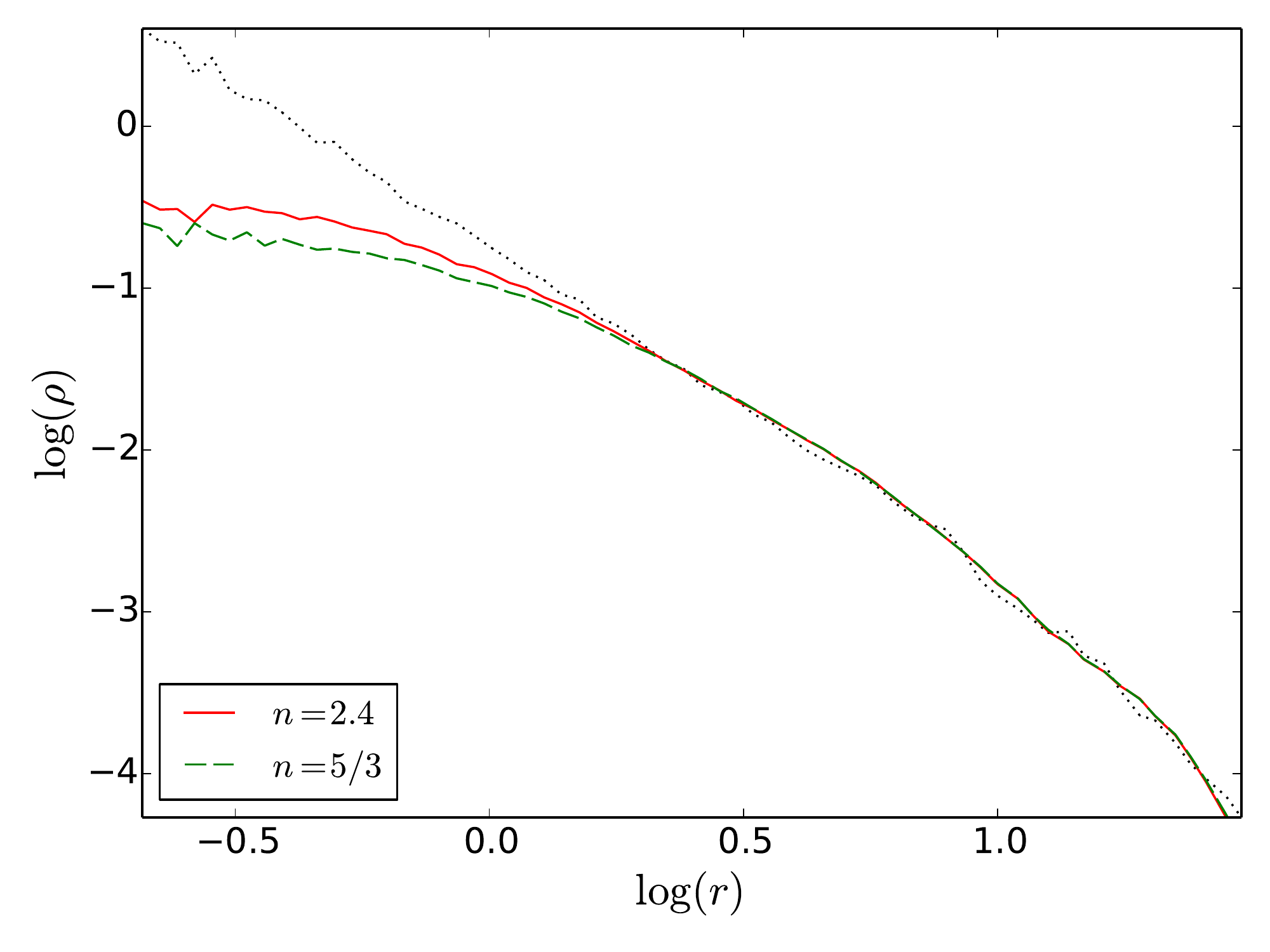}
						\caption{Dark matter density profiles after 500 Myr of perturbations for the fiducial set of parameters and for a Kolmogorov exponent $n=5/3$. Each profile has been averaged over ten random realizations. The dotted line corresponds to the initial profile; the units are as in the previous figures.} 
    \label{fig:fid-kol}
\end{figure}

Fig.~\ref{fig:fid-kol} compares the fiducial evolution of the dark matter density profile with that obtained with a Kolmogorov exponent $n=5/3$. The small power-law exponent leads to an increased flattening of the density profile within $500~\rm Myr$, which is consistent with Eq.~\ref{eq:trelax}: the relaxation time is proportional to $n$ so smaller values of the exponent result in an increased efficiency of the energy transfer to the dark matter particles. A smaller exponent also corresponds to a flatter power spectrum, i.e., to higher amplitudes at wave numbers larger than $k_{m}$. 
Nevertheless, the increased efficiency of the process remains limited due to the linear dependence of the relaxation time on the power law index $n$, 
which is likely constrained to a rather limited range of values (cf. section~\ref{section:analytics}). 
In our model, the power spectrum tilt is therefore expected to have a relatively mild influence on the final density profile.

\subsubsection{The effect of the time step}

\begin{figure}
						\centering
						\includegraphics[width=1.\columnwidth, trim = 0cm 0.3cm 0cm 0cm, clip]{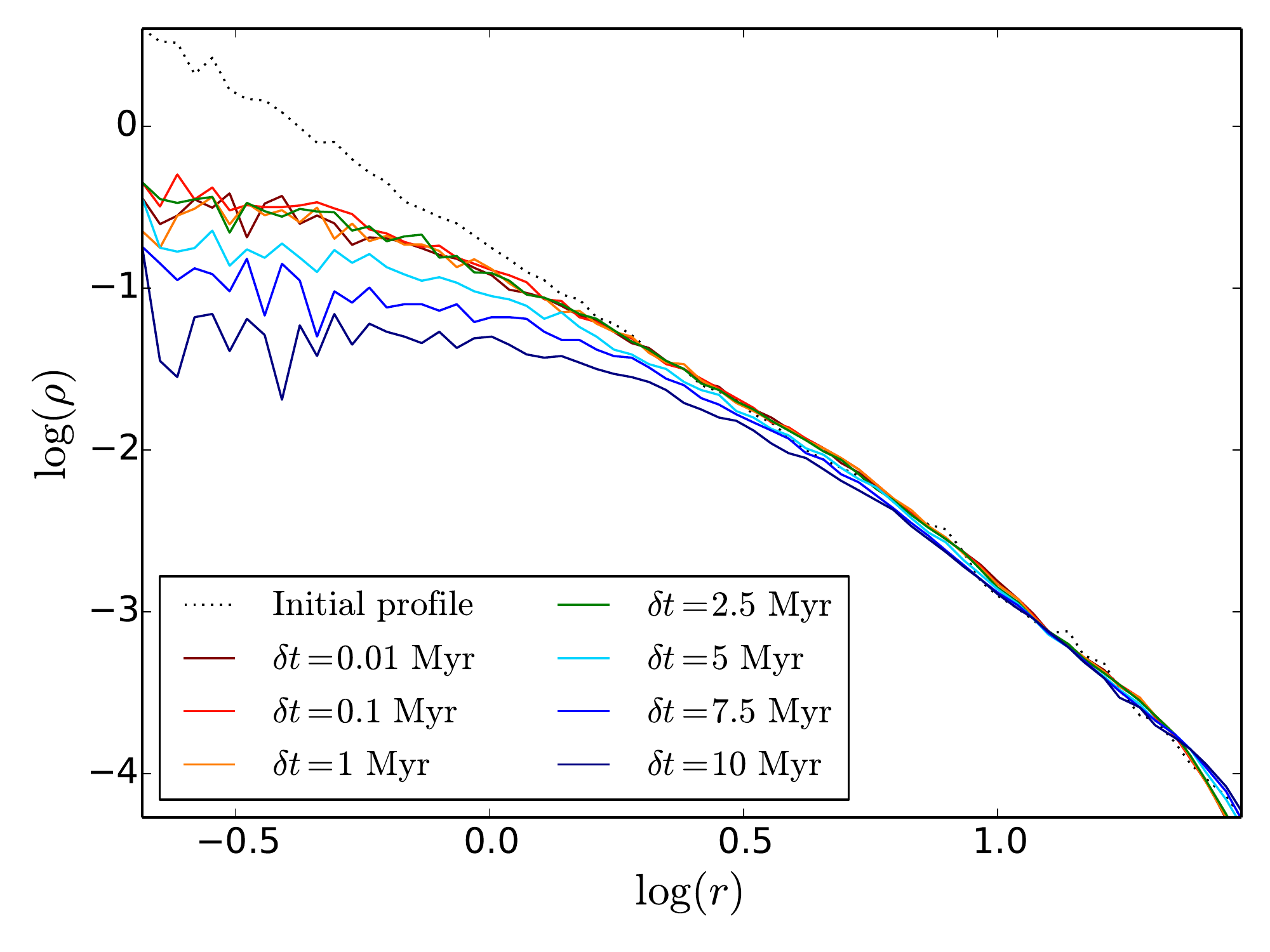}
						\caption{Dark matter density profiles after $500 \rm ~Myr$ for different values of the time step $\delta t$, in the case of density fluctuations with constant speed of turbulent flow $v_r$. Time steps $\delta t > 3 ~\rm Myr$ undersample all perturbation modes while those with  $0.03 ~\rm Myr < \delta t < 3 ~Myr$ properly sample the largest fluctuation scales close to $k_m$ but theoretically undersample the smallest ones. {The former artificially enhance the flattening while the latter lead to density profiles similar to when $\delta t = 0.01\rm~Myr$.} The dotted line corresponds to the initial profile; the units are as in the previous figures.} 
    \label{fig:timestep}
\end{figure}

To each perturbation mode between $k_{m}$ and $k_x$ corresponds a timescale, which depends on the definition of the pulsation frequency. When it is defined as $\omega (k) = v_r ~k$, the extremal timescales associated to the fiducial parameters are $T(k_{m}) = 2\pi/\omega(k_m) = 7.5$ Myr and $T(k_{x}) = 0.075 $ Myr. When it is instead defined from Larson's relation as $\omega(k) = 2\sqrt{k}$, $T(k_{m}) = 12.53$ Myr and $T(k_{x}) = 1.25$ Myr. As the smallest time scale for the fiducial simulations is $0.075~\rm Myr$, the Nyquist-Shannon sampling theorem a priori requires selecting a time step inferior to $0.03~\rm Myr$. 
However, as shown Fig.~\ref{fig:timestep}, the simulations already converge for time steps of about $1\rm ~Myr$, i.e., for time steps which only resolve the highest perturbation scales: because the amplitude of the density power spectrum decreases from $k_{m}$ to $k_{x}$ and because the small scale perturbations are swept out by the larger ones, the fluctuations are dominated by those near $k_{m}$. As simulations with $\delta t = 0.01~\rm Myr$ and $\delta t = 0.1~\rm Myr$ yield similar results, we carried out most of our simulations with the less time-consuming time step $\delta t = 0.1~\rm Myr$.

\section{Conclusion}

We have presented a theoretical model that attempts to describe, from first principles,  
halo core formation due to coupling with fluctuations in a hosted gaseous component.
Gravitational potential fluctuations leading to core formation in dark matter haloes arise 
from  density variations in the gas distribution hosted by the halo during its early evolution. 
It is then possible to understand the dynamics of core formation in terms of the statistical properties of the fluctuations in the gaseous 
density field, assumed here to be stationary in time. In particular, it is possible to derive a correlation function for the force born of the 
density fluctuations, and which affects the trajectories of the dark matter halo.
In the diffusion limit, when particles undergo random walks initiated by the persistent density fluctuations, a relaxation time, 
analogous to the two body relaxation time in $N$-body systems, can be derived. 

The framework thus described is general, and valid for any form of the density fluctuation power spectrum (provided 
the time integrals of the associated force correlation functions converge).  
However, in this initial study, we have confined ourselves to pure power-law power density fluctuation spectra, 
described by an exponent $n$ and maximal and minimal cutoff scales.     
The relaxation time does not depend on these, and depends only weakly (linearly) on $n$. 
We also assume that the gas is homogeneous on large scales, and its distribution thus 
determined by the gas mass fraction.
For a given dark matter configuration, the important parameters determining the relaxation rate are the gas 
mass fraction and the normalisation of the power spectrum of the density fluctuations.

For numerical parameters associated with a small gas-rich galaxy halo (Tables~\ref{tab:parameters-halo} and \ref{tab:parameters}), our calculations suggest that the stochastic processes discussed here can indeed lead to significant effects on the dark matter particle dynamics within a fraction of the Hubble time for a  gas mass fraction 
comparable to the universal gas baryon fraction in the region where fluctuations are significant. The magnitude of the RMS density fluctuations associated with dimensionless power spectrum need to be of order one to ten at the maximal fluctuation length scale, which is the dominant scale driving the cusp-core transformation in our model.  The mass fluctuations within radius $0.1 \rm kpc$ ($\sim 0.1 r_s$) also need to be of the same order. These appear compatible with the mass fluctuations time series  presented  by \cite{Teyssier2013}, who  do produce a core in a dwarf galaxy halo due to potential fluctuations driven by gaseous feedback and the mass fluctuations. {The fluctuation levels can be reduced if a larger gas mass fraction, representing a high level of gas condensation, is assumed.}

 This may suggest that the fluctuations leading to core formation in dark matter haloes can  be modeled as stochastic processes determined by a power spectrum and the associated dynamical effects modeled as a diffusion process; and that the relevant power spectrum may consist of a power law with maximal and minimal cutoff scales.  
As the required mass fluctuations in the central regions are large, the model can incorporate the effects 
of repeated outflows and inflows in the central region. In addition, it takes into account the effects of fragmentation
and turbulent cascades giving rise to  continuous mass and density fluctuation spectra.

Detailed comparison with full hydrodynamic simulations is left to a future study. 
However, we tested our model through $N$-body simulations in which an additional fluctuating force is imposed on the  
halo particles. This is inferred from a  Gaussian random realisations of the 
the statistical properties of the assumed gaseous field, as defined by the power spectrum of its density flcutuations. 
The simulations include the self gravity of the dark matter halo and any nonlinear coupling between 
the unperturbed halo particle trajectories and the imposed force; the diffusion limit is not assumed {\it a priori}. 
They confirm that the dynamical effects of the processes studied are independent of the process on the maximum and minimum fluctuation 
scales and only weakly dependent on the spectral index of the power-law spectrum. 
We use the self consistent field code 
of \cite{Hernquist1992},  which facilitates the isolation of effects due to collective modes. 
The results suggest that the transfer of energy  imparted from the fluctuating force on the individual 
particles, and its redistribution  through the self gravitating configuration, is greatly enhanced by the effects of collective modes born of self gravity:
the process of core formation is significantly reduced if non-radial collective modes are removed by imposing strict 
spherical symmetry on the configuration. In this latter case, the observed evolution timescales are as those inferred 
from the theoretical calculation of the relaxation time. 
In the general case, on the other hand, the process of core
formation is faster, taking place within a few hundred Myr (the aforementioned scalings   
and parameter dependence, however, remain in line with the analytical calculations). 
That stochastic fluctuations can couple 
to global modes to enhance their overall effect  is a phenomenon that has already been already observed and studied \citep{Weinberg1998}. 

Our results also suggest 
that when the time resolution of the simulations is not sufficient, the fluctuations are random sampled as white noise ($n \rightarrow 0$) and the halo 
particles feel a flat power spectrum instead of the proper one arising from the actual physical processes which are modeled, 
which leads to significant enhancement of the effect of core formation. Nevertheless, since the force fluctuations are dominated by the largest 
scales, the resulting error is not catastrophic unless even the those scales remain unresolved.

The model presented here can be used to understand how the physics leading to a particular spectrum of fluctuations
affects the dynamics of core formation in realistic simulations; and also how particular numerical implementations 
affect the process. Possible extensions include the introduction of more general power spectra and gas density 
distributions. In its present formulation,
the model predicts that the process of core formation 
primarily depends on only two parameters --- 
the normalisation of the power spectrum and the gas mass fraction inside the region where the effect of the fluctuations is important (parameters which may in fact be correlated through the star formation efficiency). 
It can  can  hence be understood  and parametrised in particularly simple terms.



\section*{Acknowledgements}

 This work benefited from the Franco-Egyptian Partenariat Hubert Curien (PHC) Imhotep and the ERC-Momentum-267399. The authors acknowledge interesting discussions with Gary Mamon, Avishai Dekel, Andrea Macci\`o, James Bullock, Joe Silk and  helpful comments from Scott Tremaine and the referee.




\bibliographystyle{mnras}
\bibliography{biblio2,bib} 




\appendix

\section{Derivation of the force correlation function}


\subsection{Expression as an integral}

	The Wiener-Khinchin theorem enables to write the force auto-correlation function as the inverse Fourier transform of the force power spectrum so that, when assuming isotropy, 
		\begin{equation}
		\langle {\bf F}(0) . {\bf F}(r) \rangle = \frac{1}{\left(2\pi\right)^3} \int^\infty_0 \mathcal{P}_F({ k})~4\pi k^2~\frac{\sin~kr}{kr}~dk.
		\end{equation} 
    Given Eq.~(\ref{eq:phikk}) and (\ref{eq:pfk}), 
	    \begin{equation}
		    \mathcal{P}_F(k) = V ~\left(4\pi G \rho_0 \right)^2 ~k^{-2} ~\langle|\delta_{{k}}|^2\rangle
	    \end{equation}
	so for power law density fluctuations bounded by $k_m$ and $k_x$ as defined in section~\ref{section:forcecor}, 
		\begin{equation}
		\label{eq:app/egypt-analytics/fsin}
		\langle {\bf F}(0) . {\bf F}(r) \rangle = \frac{D}{r} \int^{k_x}_{k_m} \frac{\sin~kr}{k^{n+1}}~dk
		\end{equation}
	with $D=8 \left(G\rho_0\right)^2 C d^3 $. 
	
\subsection{In terms of incomplete Gamma functions}
%


		The upper incomplete Gamma function is defined as
		\begin{equation}
		\Gamma(s,x) = \int_x^\infty t^{s-1} e^{-t} dt.
		\end{equation}
		It can be expressed as power series, and as such, developed into a holomorphic function of complex variables with the same properties. 
		%
		%
%
		The force auto-correlation function can be expressed in terms of incomplete Gamma functions extended for complex variables:
		\begin{equation}
		\label{eq:F0FR}
		\begin{array}{ll}
		\displaystyle\langle  {\bf F}(0) . {\bf F}(r) \rangle &= \displaystyle \frac{D}{r} \frac{1}{2i} \int_{k_{m}}^{k_{x}} \frac{e^{ikr}-e^{-ikr}}{k^{n+1}}~dk\\
			&= \displaystyle \frac{D}{r} \frac{1}{2i} \left(ir\right)^{n}\int_{ik_{m}r}^{ik_{x}r} \frac{e^{x}-e^{-x}}{x^{n+1}}~dx\\
		\displaystyle \langle {\bf F}(0) . {\bf F}(r) \rangle &= \displaystyle \frac{D \left(ir\right)^{n-1}}{2} \Big( \Gamma(-n,ik_{x}r)-\Gamma(-n,ik_{m}r)\Big)+C.C.\\
		\end{array}
	    \end{equation}

\subsection{Asymptotic behavior}

		Given that 
				\begin{equation}
				\label{eq:asymptgamma}
				\Gamma(s,x) \underset{|x|\rightarrow+\infty}{\sim} x^{s-1} e^{-x}, 
				\end{equation}
		the asymptotic behavior of the force correlation function when $k_{x}r \gg k_{m}r \gg 1$ and $n+1>0$ is 
		\begin{equation}
		\begin{array}{ll}
		\displaystyle \langle {\bf F}(0) . {\bf F}(r) \rangle
			&\sim \displaystyle -\frac{D}{2 r^2} \left[ k_{x}^{-n-1} e^{-ik_{x}r}-k_{m}^{-n-1}e^{-ik_{m}r}\right]+C.C.\\
			&\sim \displaystyle \frac{D}{r^2} \left[ k_{m}^{-n-1} \cos(k_{m}r)-k_{x}^{-n-1} \cos(k_{x}r)\right]\\
		\displaystyle \langle {\bf F}(0) . {\bf F}(r) \rangle
			&\sim \displaystyle \frac{D}{r^2}\frac{1}{k_{m}^{n+1}} \cos(k_{m}r). 
		\end{array}
		\end{equation} 

\subsection{An estimate of the force}

The value of $\langle F(0)^2 \rangle$ can be used as an estimate of the square of the force.  Eq.~\ref{eq:app/egypt-analytics/fsin} yields
		\begin{equation}
		\langle F(0)^2\rangle 
			= D \int_{k_{m}}^{k_{x}} \frac{1}{k^{n}}  ~dk
			= \frac{D}{n-1} \bigg(k_{m}^{-n+1}-k_{x}^{-n+1}\bigg)
		\end{equation} 
		so that
		\begin{equation}
		\langle F(0)^2\rangle = \frac{8\left(G\rho_0\right)^2 \langle \delta^2_{k_m} \rangle ~d^3}{n-1} ~k_{m} ~ \left(1-\left(\frac{k_{m}}{k_{x}}\right)^{n-1}\right). 
		\end{equation} 
		%


\section{Velocity variance}

\subsection{Expression from the equation of motion}
\label{section:appendixB1}

Considering the effect of the random perturbation force in direction $i$ during a time $T$, the equation of motion
leads to
		\begin{equation}
		\frac{dx_i}{dt} = v_{0i} +\int_0^T F_i(\tau) d\tau, 
		\end{equation}
where $v_{0i}$ is the initial velocity in direction $i$. The velocity variance is obtained by averaging this equation:
		\begin{equation}
		\displaystyle \langle (\Delta v_i)^2 \rangle 
			= \displaystyle \langle \left(\frac{dx_i}{dt}-v_{0i}\right)^2\rangle 
			= \displaystyle \int_0^T \int_0^T  \langle F_i(\tau) F_i(\tau^\prime) \rangle~ d\tau d\tau^\prime.
		\end{equation}
The integrand is symmetrical in $\tau$, $\tau^\prime$ and the integration domain correspond to a square of length $T$ in the corresponding plane. We can thus replace the integral over the square by twice the integral over the triangle defined by $0<\tau<T$ and $\tau<\tau^\prime<T$ so that
		\begin{equation}
		\displaystyle \langle (\Delta v_i)^2 \rangle 
		=\displaystyle 2 \int_0^T  d\tau \int_\tau^T  d\tau^\prime ~\langle F_i(\tau) F_i(\tau^\prime) \rangle, 
		\end{equation}	
which can be rewritten as
		\begin{equation}
		\displaystyle \langle (\Delta v_i)^2 \rangle =\displaystyle 2 \int_0^T  dt \int_0^{T-t}  d\tau ~\langle F_i(\tau) F_i(\tau+t) \rangle.
		\end{equation}
		The perturbations being stationary,  $\langle F_i(\tau) F_i(\tau+t) \rangle = \langle F_i(0) F_i(t) \rangle$, the expression simplifies to
		\begin{equation}
		\displaystyle \langle (\Delta v_i)^2 \rangle 
		=\displaystyle 2 \int_0^T \left(T-t\right) \langle F_i(0) F_i(t) \rangle~ dt.
		\end{equation}
		and the total velocity variance is given by
		\begin{equation}
		\label{eq:app/egypt-analytics/var}
		\langle (\Delta v)^2 \rangle =\displaystyle 2 \int_0^T \left(T-t\right) \langle {\bf F}(0) . {\bf F}(t) \rangle~ dt.
		\end{equation}
		As indicated in section \ref{section:velvar}, this quantity can also be expressed in terms of the spatial correlation function $\langle {\bf F}(0) . {\bf F}(r) \rangle$ by introducing a velocity $v_r$ corresponding to the movement of the fluctuating gaseous field (Eq. \ref{eq:intF0FR}). 

	\subsection{Explicit expression of the velocity variance}
	\label{section:app/egypt-analytics/variance/explicit}
	
	The expression of the force auto-correlation function obtained in Eq.~(\ref{eq:F0FR}) can be separated in two analogous components, one depending on $k_x$ and the other on $k_m$. The one depending on $k_x$ can be developed as
	 	\begin{equation}
			\langle {\bf F}(0) . {\bf F}(r) \rangle_{k_x}
			= \frac{D}{2} k_{x}^{-n+1} \left( \left(ik_{x}r\right)^{n-1} \Gamma(-n,ik_{x}r)+ C.C.\right)
		\end{equation}
	and results for the velocity variance (as expressed by Eq. \ref{eq:intF0FR}) in a component 
		\begin{equation}
			\langle (\Delta v)^2 \rangle_{k_x}
			=  \frac{D}{v_r^2} k_{x}^{-n+1} \Bigg( -\frac{iR}{k_x} \frac{I_{1}(k_xR)}{n} + \frac{1}{k_x^2} \frac{I_{2}(k_xR)}{n+1}\Bigg)
		\end{equation}
	with
		\begin{equation}
			\frac{I_{1} (k_xR)}{n}=  \int_{-ik_x R}^{ik_{x}R} x^{n-1} \Gamma(-n,x)~dx
		\end{equation}
	and
		\begin{equation}
			\frac{I_{2} (k_xR)}{n+1}=  \int_0^{ik_{x}R} x^{n} \Gamma(-n,x)~dx+ C.C.
		\end{equation}	
	Given that 
			\begin{equation}
			\label{eq:app/egypt-analytics/integralGamma}
			\int x^{b-1} ~\Gamma (s,x) ~dx = \frac{1}{b}\left[ x^b \Gamma (s,x) - \Gamma (s+b,x)\right], 
			\end{equation}
	integrating by parts yields 
		\begin{equation}
			I_{1} (k_xR) =   \left(ik_{x}R\right)^n \Gamma(-n,ik_{x}R) - \Gamma(0,ik_{x}R) - C.C.
		\end{equation}
	and
		\begin{equation}
			I_{2} (k_xR) = \left(ik_{x}R\right)^{n+1} \Gamma(-n,ik_{x}R) 
						-\Gamma(1,ik_{x}R)+\Gamma(1,0) +C.C. 
		\end{equation}		
	Further noticing that 
		\begin{equation}
			\Gamma(0,ik_{x}R)-C.C. = 2 i ~{\rm Si}(k_x R),
		\end{equation}
	where $\displaystyle {\rm Si}(X)\equiv \int_{0}^{X} \frac{\sin t}{t}~dt$ is the Sine integral function, and that 
		\begin{equation}
			\Gamma(1,0)-\Gamma(1,ik_{x}R)+C.C. = -2 \big(\cos (k_x R)-1\big), 
		\end{equation}	
	we obtain 
		\begin{equation}
			I_{1}(k_xR) =    \Big(\left(ik_{x}R\right)^n \Gamma(-n,ik_{x}R) - C.C.\Big)  -{2i}~{\rm Si}(k_{x}R)
		\end{equation}	
	and
		\begin{equation}
			I_{2} (k_xR) = \Big( \left(ik_{x}R\right)^{n+1} \Gamma(-n,ik_{x}R) + C.C. \Big) -{2} \Big(\cos(k_{x}R)-1\Big).
		\end{equation}			
	Hence, 
		\begin{equation}
		\langle (\Delta v)^2 \rangle_{k_x} = - \frac{D R}{v_r^2 k_x^{n}}  \left(\frac{2}{n} {\rm Si}\left(k_{x} R\right)+T_1(k_{x} R)+T_2(k_{x} R)\right)
		\end{equation}
	while the component depending on $k_m$ similarly yields
		\begin{equation}
		\langle (\Delta v)^2 \rangle_{k_m} = \frac{D R}{v_r^2 k_m^{n}}  \left(\frac{2}{n} {\rm Si}\left(k_{m} R\right)+T_1(k_{m} R)+T_2(k_{m} R)\right), 
		\end{equation}	
	where the functions $T_1$ and $T_2$ have been defined in Eq.~(\ref{eq:T1}) and (\ref{eq:T2}). The total velocity variance is simply $\langle (\Delta v)^2 \rangle = \langle (\Delta v)^2 \rangle_{k_x} + \langle (\Delta v)^2 \rangle_{k_m}$, which simplifies to its second term when $k_x \gg k_m$. 
	
\subsection{Asymptotic behavior}

Eq.~(\ref{eq:asymptgamma}) results in 
	\begin{equation}
		T_1(k_x R) \sim \left(\frac{1}{n} - \frac{1}{n+1} \right) \frac{2}{k_x R} \cos(k_x R)
	\end{equation}
when $k_x R\gg 1$ so all the terms in $T_1$ and $T_2$ go to zero when $k_x R \gg 1$ and $k_m R \gg 1$. In this limit and when $k_x \gg k_m$, we consequently have
		\begin{equation}
		\langle (\Delta v)^2 \rangle 
		\sim \frac{\pi R D}{n v_r^2} \frac{1}{k_{m}^n}. 
		\end{equation}


\bsp	
\label{lastpage}
\end{document}